\documentclass[12pt,preprint2]{emulateapj}

\newcommand{\rmg}{{\cal R}_{{\rm mg}}}
\newcommand{\rmgv}{{\cal R}_{{\rm mg}_{V}}}
\newcommand{\msun}{M_{\odot}}
\newcommand{\lsun}{L_{\odot}}

\slugcomment{Accepted for publication in the Astrophysical Journal}

\shorttitle{The Luminosity Dependence of the Galaxy Merger Rate} 
\shortauthors{D. R. Patton \& J. E. Atfield}

\begin{document}

\title{The Luminosity Dependence of the Galaxy Merger Rate}

\author{D. R. Patton\altaffilmark{1} and J. E. Atfield\altaffilmark{1}
}

\altaffiltext{1}{Department of Physics \& Astronomy, Trent University, 
1600 West Bank Drive, Peterborough, ON, K9J 7B8, Canada; 
dpatton@trentu.ca, julianatfield@trentu.ca}

\begin{abstract}

We measure the number of companions per galaxy ($N_c$) as a function of 
$r$-band absolute magnitude for both the Sloan Digital Sky Survey 
and the \citet{croton06} 
semi-analytic catalog applied to the Millennium Run simulation.  
For close pairs with projected separations of 5-20 $h^{-1}$ kpc, 
velocity differences less than 500 km s$^{-1}$, and luminosity ratios between
1:2 and 2:1, we find good agreement between the observations and 
simulations, with $N_c$ consistently 
close to 0.02 over the range $-22 < M_r < -18$.  For larger pair separations, 
$N_c(M_r)$ instead becomes increasingly steep towards the faint end, 
implying that luminosity-dependent clustering plays an important role 
on small scales.
Using the simulations to assess and correct for projection effects, 
we infer that 
the {\em real-space} $N_c(M_r)$ for close pairs peaks at about $M^*$, and 
declines by at least a factor of two as $M_r$ becomes fainter.  
Conversely, 
by measuring the number density of close companions, we estimate that 
at least $90\%$ of all major mergers occur between galaxies which are 
fainter than $L^*$.  Finally, measurements of the luminosity 
density of close companions indicate that $L^*$ galaxies likely dominate in 
terms of the overall importance of major mergers in the evolution of 
galaxy populations at low redshift.

\end{abstract}

\keywords{galaxies: evolution, galaxies: interactions, 
surveys, galaxies: statistics}

\section{Introduction}\label{intro}

Galaxy mergers can produce dramatic changes in the morphological, nuclear 
and star forming properties of galaxies over relatively short timespans.  
As a result, mergers have been invoked to explain a number of 
aspects of galaxy evolution.
In recent years, large redshift surveys have paved the way for 
systematic observational studies of 
candidate mergers, while theoretical modelling of structure formation 
has produced key advances in our understanding of the role of 
merging in a cosmological context.  
In general, these efforts have focussed on two key aspects of merging: 
(1) the effects of merging on the constituent galaxies, and 
(2) the frequency with which merging occurs, 
as described by the merger rate and related quantities.

An increasingly popular method of identifying candidate merging systems 
is the use of close galaxy pairs, which are the precursors to mergers.  
With careful choices of close pair criteria, and correction for 
projection effects (ie., contamination by non-merging pairs), 
the frequency of close pairs should correlate with the merger rate.
Recent studies using cosmological simulations support this idea, 
demonstrating that most close pairs merge on relatively short 
timescales \citep{kitzbichler08}.  In addition, 
the properties of paired galaxies can provide insight into the 
nature of merging galaxies both before and during the encounter.  
Galaxies in close pairs have higher asymmetries than galaxies in wider 
separation pairs or the field \citep{hernandez05,patton05,depropris07}, 
confirming that interactions and mergers are prevalent in these systems.
Star formation is enhanced in close pairs at low redshift
\citep{2dfpairs,2dfpairsb,nikolic04,patton05,alonso06,geller06,barton07,owers07,smith07,woods07,ellison08,li08}, 
implying that star formation has been triggered by 
mergers or interactions.  Differences in metallicities between paired 
and field galaxies are consistent with a scenario in which  
interactions funnel gas to the central regions of galaxies involved 
in these close encounters \citep{kewley06,ellison08}.
Most ultraluminous infrared galaxies (ULIRGs) originate from major mergers of 
gas rich galaxies \citep{dasyra06a,dasyra06b}, while approximately 
half of the luminous infrared galaxies (LIRGs) at low redshift 
appear to be undergoing interactions or mergers \citep{wang06}.

Using close galaxy pairs and/or galaxy asymmetries as indicators of 
imminent or recent mergers, the merger rate and its evolution has now 
been measured using a number of large redshift surveys
\citep{carlberg00,lefevre00,cnoc2mr,conselice03,bundy04,lin04,bell06a,kartaltepe07,kampczyk07,conselice08,hsieh08,lin08,lotz08,rawat08,ryan08}.  
Evolution estimates range from roughly 
$(1+z)^{0.5}$ to $(1+z)^3$, implying widely differing scenarios at 
$z \sim 1$ and above.  At least some of these discrepancies result from 
the use of different pair criteria.  
For example, simulations and semi-analytical models of galaxy formation 
indicate that the merger rate and its evolution depends on factors 
such as galaxy mass, pair mass ratio and environment
\citep{khochfar01,berrier06,maller06,cox08,guo08,kitzbichler08}. 

In order to better understand the role of merging, and to reconcile 
merger rate measurements from disparate samples, we need to 
assess how the frequency and nature of merging 
depends on factors such as 
environment, 
mass ratio (ie., major versus minor mergers), 
properties of the progenitor galaxies 
(e.g., dry mergers versus gas-rich mergers), 
and the mass (or luminosity) of the merging galaxies or merger remnants 
(e.g., formation of massive galaxies versus $L^*$ galaxies).  
Significant observational progress has been made in all of these areas 
in recent years.  
Galaxy groups appear to be an ideal environment for mergers
\citep{goto05,brough06,miles06,robotham06,zandivarez06,coziol07,nolan07},
though the infall regions of 
clusters \citep{vandokkum99,tran05,moss06} and the low density 
field \citep{barton07} are important too.
Induced star formation appears to be strongest in major mergers, or
in the lower luminosity (or mass) members of minor mergers 
\citep{woods06,woods07,ellison08}.
Dry mergers have been invoked to explain the assembly of massive 
elliptical galaxies since $z \sim 1$ \citep{vandokkum05,bell06b,naab06}, 
though this process likely cannot explain all recently formed 
early type galaxies \citep{cox06,brown07,bundy07,scarlata07}.

In this paper, we investigate the luminosity dependence of the 
merger rate at low redshift, using close galaxy pairs 
in the Sloan Digital Sky Survey \citep{york00}, and in 
the \citet{croton06} semi-analytic galaxy catalogs derived from the 
Millennium Run simulation \citep{springel05}.  
Both of these samples are large enough that, in addition to being
able to measure close pair statistics as a function of absolute magnitude, 
we also have the luxury of being 
very selective in how we choose our close pairs.  In particular, 
we require all of our pairs to have spectroscopic redshifts for both members, 
projected separations less than 20 $h^{-1}$ kpc, and relative velocities 
less than 500 km s$^{-1}$.  In addition, we require our companion 
sample to be volume 
limited for all luminosity ratios between 1:2 and 2:1, thereby providing 
a cleaner match to the major merger candidates we seek to identify.
Moreover, our measurements are carried out in the $r$-band, yielding 
absolute magnitudes that are a better proxy for stellar mass than 
those at shorter wavelengths.  This is beneficial for close pair studies, 
since merger-induced star formation is likely to affect the luminosities
of galaxies in pairs more than normal (isolated) galaxies.

We begin by describing the creation of our 
SDSS spectroscopic and photometric samples in \S~\ref{data}.  
Section \ref{ncsdss} outlines the calculation of the number of 
close companions per galaxy ($N_c$) for SDSS, 
including corrections for spectroscopic incompleteness.  
We make a direct comparison 
with the Millennium Run simulation in \S~\ref{ncmill}, and 
derive real space pair statistics for both SDSS and Millennium 
in \S~\ref{realspace}.  With additional assumptions, we then 
relate these close pair statistics to the merger 
rate in \S~\ref{mrate}.
We summarize our  
conclusions  in \S~\ref{conclusions}.
Throughout this study, we adopt cosmological parameters of 
$\Omega_m = 0.3$, $\Omega_\Lambda = 0.7$, and 
$H_0 = 100~h$ km s$^{-1}$ Mpc$^{-1}$.  For brevity, 
we express all absolute magnitudes as $M_r$ instead of $M_r - 5 \log (h)$.

\section{Data} \label{data}
The fifth data release (DR5) of the Sloan Digital Sky Survey (hereafter SDSS)
consists of $ugriz$ imaging spanning over 8000 deg$^2$, along with 
spectra of about one million galaxies, quasars, and stars within a
5713 deg$^2$ subset of the imaging area \citep{DR5}.  In this study, 
we use the main galaxy sample, as described by \citet{strauss02}.  
Unlike the luminous red galaxy sample of \citet{eisenstein01}, 
this sample is designed to be independent of galaxy luminosity and 
Hubble type, and therefore probes a representative sample of galaxies, 
including both gas-rich and gas-poor (``dry'') merger candidates. 

Our goal is to carry out a census of galaxies with close physical companions.
We restrict our observational sample to close galaxy pairs for which redshifts
are available for both galaxies.  This reduces contamination due to 
non-merging systems, and allows one to measure rest-frame properties 
of the pairs and their member galaxies.  However, the SDSS is not a complete 
spectroscopic sample, and in fact the minimum fiber separation of 
$55\arcsec$ biases the sample against the close pairs we are interested in.
Fortunately, it is possible to measure and correct for spectroscopic 
incompleteness by using the photometry of all galaxies, regardless of
whether or not spectra were obtained.  To this end, we now describe the 
creation of both spectroscopic and photometric catalogs of galaxies 
drawn from the main galaxy sample.

\subsection{Spectroscopic Catalog}

The SDSS main galaxy sample consists of galaxies 
with $r$-band Petrosian magnitudes of 
$m_r \leq 17.77$ \citep{strauss02}, after correction for 
Galactic extinction.  However, regions covered by the first 
data release of the survey had limiting apparent magnitudes ranging from 
17.5 to 17.77 \citep{abazajian03}.  In addition, the sample becomes 
incomplete at the bright end, where automated deblending of large galaxies  
becomes unreliable \citep{strauss02}, and introduces 
concerns about single galaxies mistakenly being classified as close pairs.
To ensure a complete and reliable sample, 
we therefore begin by restricting our catalogs to the range 
$14.5 \leq m_r \leq 17.5$.  We also ensure that every galaxy has a 
measured spectroscopic redshift; moreover, we require the SDSS
``zConf'' parameter to be at least 0.7, thereby ensuring that 
the confidence in each redshift measurement is at least 70\% (in most 
cases, it is much higher).  
Using these criteria, we create a preliminary spectroscopic sample by 
querying the SDSS 
online database\footnote{http://casjobs.sdss.org/CasJobs/}. 

For every galaxy in the spectroscopic sample, we measure 
the $r$-band absolute magnitude at redshift $z$, as given by
\begin{equation} \label{eqnmr}
M_r = m_r - 5 \log d_L(z) - 25 - k_r - E(z),
\end{equation}
where $m_r$ is the extinction-corrected Petrosian $r$-band magnitude, 
$d_L$ is the luminosity distance, and 
$k_r$ and $E(z)$ are the $k$-corrections and passive stellar evolution
corrections, respectively.  We measure $k$-corrections using the 
SDSS $ugriz$ photometry, employing 
the kcorrect software (version v1\_1\_4) of \citet{blanton07}.  
Following \citet{cnoc2mr}, we parameterize the evolution correction as 
$E(z) = -Qz$, where $Q$ is determined from measurements 
of the galaxy luminosity function (LF) and 
$z$ is the redshift \citep{cnoc2lf}.  
We adopt $Q = 1.8$ as an average of 
the SDSS $^{0.1}r$ and $^{0.1}g$ $Q$ measurements 
of \citet{blanton03a}, since rest-frame $r$ lies between these two passbands 
(this agrees well with the evolutionary correction of \citet{tegmark04}).
Given the relatively low redshift of our sample, however, 
this evolution correction is small, and removing it entirely (ie., $Q = 0$) 
does not significantly change any of the conclusions in this paper.

\subsection{Photometric Catalog}

In order to measure and correct for incompleteness in our spectroscopic 
catalog, we also create a photometric catalog, in which galaxies satisfy 
the same flux limits as the spectroscopic catalog, but are not required
to have a measured redshift.
The sky area covered by DR5 imaging  
is about 40\% larger than the area with spectroscopic coverage \citep{DR5}; 
to match the footprints of our spectroscopic and photometric catalogs, 
we therefore also require every galaxy in both samples to have at least 
one galaxy with a spectrum within 12 arcminutes of its position 
(excluding itself).

\subsection{Spectroscopic Incompleteness}\label{wz}

By comparing our resulting photometric and spectroscopic catalogs, 
we find that the average spectroscopic completeness is 88\%.  While 
spectroscopic target selection was designed to provide uniform 
coverage \citep{blanton03b}, 
the completeness does vary considerably from one part of the sky
to another, with the completeness in some regions falling well below
the mean, while in other regions (particularly those covered by more than one 
SDSS plate), the completeness is considerably higher.
\citet{cnoc2mr} demonstrated that the observed number of companions per
galaxy scales with the spectroscopic completeness.  While this 
bias can be corrected for, one can also minimize its 
impact by excluding regions with low spectroscopic completeness.  
Such an exclusion will also remove galaxies which lie close enough to the 
survey boundaries that close companions will be missed.
With this in mind, and with a desire to measure and correct for the 
remaining spectroscopic incompleteness, we compute a measure of 
local spectroscopic completeness for every galaxy in our spectroscopic
and photometric catalogs. 
We consider an area around each galaxy within an outer radius 
of one degree, and an inner radius of $55\arcsec$ (the latter corresponds
to the minimum fiber separation).  
After counting the number of enclosed
galaxies in the spectroscopic and photometric samples, we take the 
ratio of these two numbers, denoting this quantity $f_s$.
We then require every galaxy in our catalogs to have $f_s > 0.75$.  
This restriction excludes 4.7\% of the galaxies in the spectroscopic catalog.  
The main conclusions of this paper are unchanged if we instead use 
$f_s > 0.7$ or $f_s > 0.8$ (excluding 2.5\% and 9.5\% of the 
spectroscopic catalog respectively).

\section{SDSS Close Pair Statistics}\label{ncsdss}

\subsection{Methodology}\label{methodology}

We now proceed to identify close galaxy pairs from our spectroscopic 
sample.  We measure three key quantities for each galaxy pair:  
projected physical separation $r_p$, rest-frame relative velocity 
$\Delta v$, and absolute magnitude difference $\Delta M_r$.
Following \citet{ssrs2mr,cnoc2mr}, we define a ``close pair'' to 
have $5 < r_p < 20~h^{-1}$ kpc and $\Delta v < 500$ km s$^{-1}$.  
On order half 
of the pairs satisfying these criteria are known to exhibit 
morphological signs of interactions, based on visual classification 
\citep{ssrs2mr} and quantitative measures of 
asymmetry \citep{patton05,depropris07}.
In addition, in order to focus on major merger 
candidates, we require that $|\Delta M_r| \leq 0.753$, ensuring 
a pair luminosity ratio between 1:2 and 2:1.  This criterion is preferable 
to the more common approach of selecting both host and companion galaxies from
the same fixed range in absolute magnitude 
(e.g., \citet{ssrs2mr,cnoc2mr,lin04,depropris05}), since that approach 
includes a wider range in luminosity ratios, and becomes increasingly 
incomplete at luminosity ratios significantly different from 1:1.

\subsection{Choosing Potential Host and Companion Galaxies}
\label{hostcomp}

\begin{figure}
\plotone{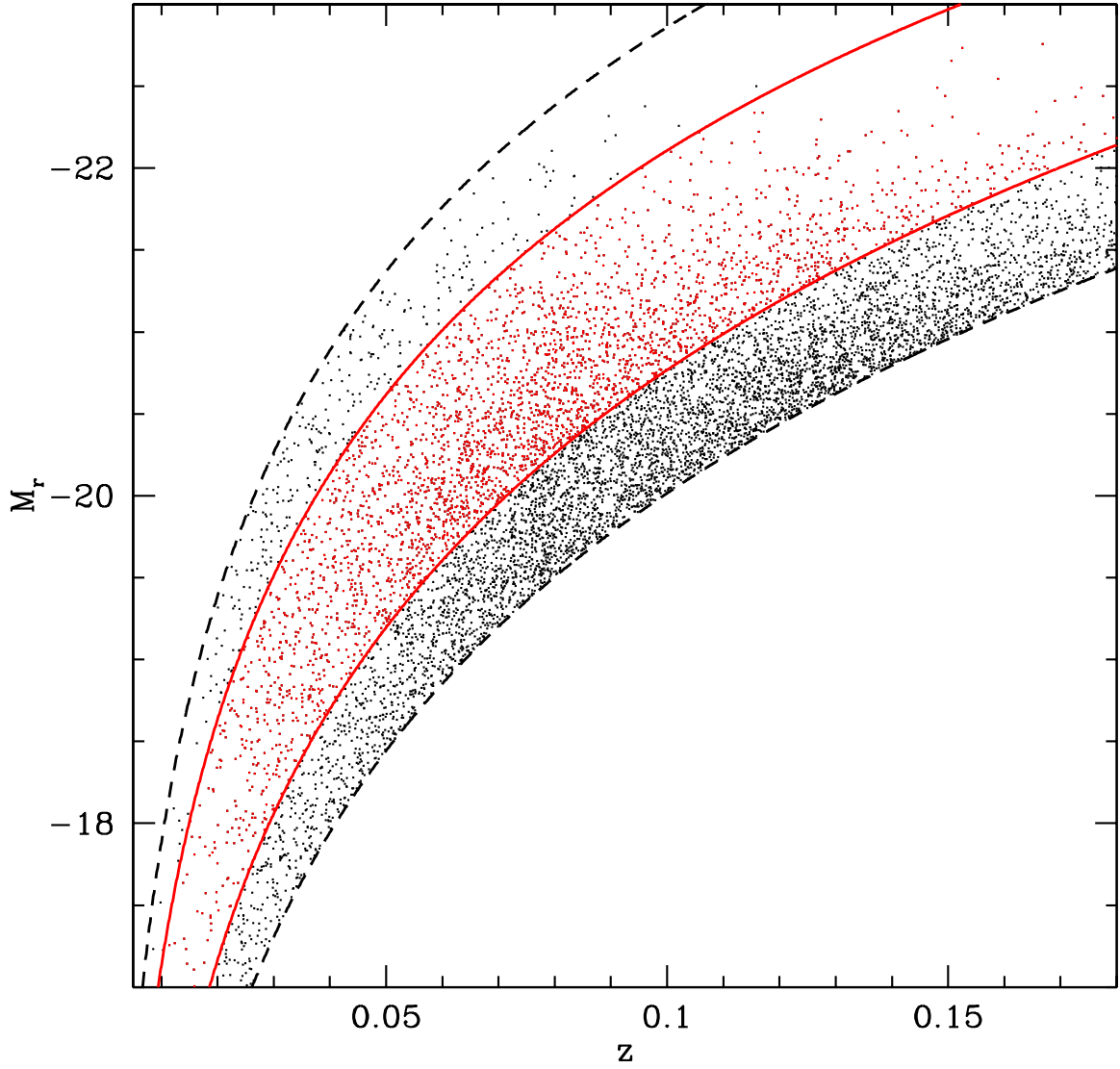}
\caption{Absolute magnitude is plotted versus redshift for 10,000 galaxies 
selected at random from our SDSS spectroscopic catalog of about 337,000 
galaxies.  
The dashed (black) lines enclose all potential companion galaxies; galaxies 
within this region have $14.5 < m_r < 17.5$.
The solid (red) lines enclose all potential host galaxies; galaxies within 
this region differ by at least 0.753 magnitudes from the dashed lines.  
For these potential host galaxies, all companions within a luminosity ratio of 
1:2 are therefore detectable.
\label{absmag}}
\end{figure}

Figure~\ref{absmag} contains a plot of absolute magnitude 
versus redshift for 10,000 galaxies selected at random 
from our spectroscopic sample of about 337,000 galaxies.  
We begin our sample selection by 
measuring the bright and faint limits 
in absolute magnitude as a function 
of redshift within which galaxies of all spectral types will have 
$14.5 \leq m_r \leq 17.5$.  These are shown by the upper and lower 
dashed (black) curves in Figure~\ref{absmag}, and are computed using 
estimates of the minimum and maximum k-corrections respectively.
We denote all galaxies lying within these limits as 
{\it potential companion galaxies}.  

Within this sample, we then identify the subset of galaxies
for which all companions (with $|\Delta M_r| \leq 0.753$) are detectable.
These galaxies, which we refer to as {\it potential host galaxies}, 
are contained within the two solid (red) curves 
in Figure~\ref{absmag}.  Aside from spectroscopic incompleteness
(which we correct for statistically), this provides a sample of host 
galaxies for which we can carry out a volume limited search for companions.
It follows that our sample will contain close pairs in which either one or 
both galaxies fall into the sample of potential hosts.

We wish to compute pair statistics as a function of $M_r$, over as 
large a range as feasible.  However, in order to ensure completeness, 
we must impose some restrictions on the range of host galaxy luminosities.
At the bright end, Figure~\ref{absmag} indicates that our sample contains 
relatively few galaxies close to $M_r \sim -23$.  
Given the need to detect companions which are 
0.753 magnitudes brighter than any potential host galaxy, we therefore
impose a minimum $M_r = -22$.  
At the faint end, the key issue is to decide on a reasonable minimum 
redshift to use, as this will dictate a maximum allowable absolute 
magnitude.  Galaxies at the lowest redshifts have the least certain 
absolute magnitudes, due to the increased influence of peculiar velocities
on the observed redshift.  In addition, such nearby galaxies are typically 
the most challenging to obtain accurate photometry for, since the 
SDSS deblending algorithm tends to break down more often for 
galaxies with large apparent sizes.  With these considerations in mind, 
we impose a conservative maximum $M_r$ of $-18$ on our sample of 
potential host galaxies, corresponding to a minimum redshift of 0.015.
Companions are permitted to lie at slightly lower redshifts due to the 
allowed velocity difference of 500 km s$^{-1}$.

With these criteria, we find a total of 477 host galaxies which have 
at least one close companion.  
In order to ensure that the SDSS 
pipeline has been successful in detecting real galaxy pairs, we 
visually inspect the SDSS images of every detected host galaxy.  
The contamination was
found to be very small, with only 0.8\% of the host galaxies being
spurious.  The affected systems consist of 
one edge-on disk, and one ring galaxy: in both cases, the SDSS pipeline 
mistakenly detected two galaxies.   We therefore remove these galaxies 
from our sample, leaving 473 host galaxies in the 
range $-22 < M_r < -18$.  The basic properties of these galaxies are 
listed in Table~\ref{tabhost} (this table contains a subset of the 
table, which is to be published in its entirety in the electronic
version of the journal). 

A significant fraction
of these galaxies exhibit morphological signs of interactions, though 
we defer a more rigorous structural analysis to a future paper.
We note that this sample is more than twice as
large as the Millennium Galaxy Catalogue sample of \citet{depropris05}, 
despite our more rigid requirement that the sample
be volume-limited for close companions which are 
major merger candidates ($|\Delta M_r < 0.753|$).

\subsection{Small Scale Spectroscopic Incompleteness}\label{wtheta}

In Section~\ref{wz}, we described our algorithm for measuring the local 
spectroscopic completeness for every galaxy.  This provides a measure of
completeness on scales of on order half a degree.  However, 
constraints on fiber placement require angular separations of  
at least $55\arcsec$ between any two targets assigned to the same plate 
\citep{strauss02}.  
As a result, spectroscopic completeness 
drops sharply at smaller angular pair separations.  At a redshift of 0.1, 
$55\arcsec$ corresponds to a projected separation of 71 $h^{-1}$kpc, 
meaning that 
most of the close pairs of interest in this study ($r_p < 20~h^{-1}$ kpc) 
are found at these small angular separations.  
Fortunately, most plates contain regions of overlap with adjacent plates, 
and some regions are observed using two or more plates.  
As a result, the spectroscopic 
sample contains enough close angular pairs that we are able to model the 
incompleteness and correct for it, 
following the method of \citet{cnoc2mr}.  

\begin{figure}
\plotone{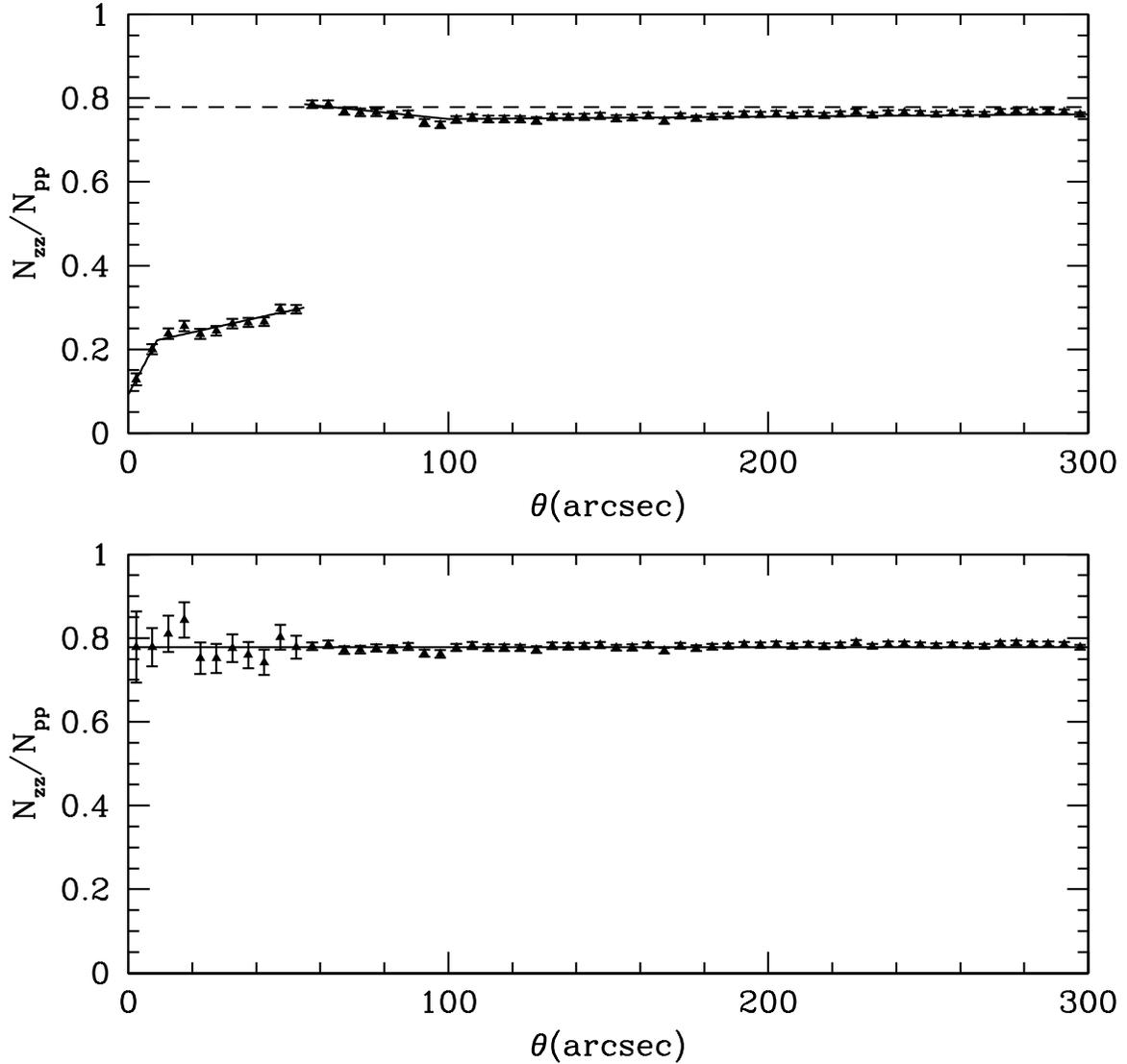}
\caption{The ratio of spectroscopic to photometric pairs ($N_{zz}/N_{pp}$) 
is plotted versus angular separation ($\theta$), with error bars 
computed using the Jackknife technique.  (a) The upper panel 
shows the data before corrective weights have been applied, and a substantial 
deficit of spectroscopic pairs is seen at small separations ($\theta < 
55\arcsec$).  At larger separations, $N_{zz}/N_{pp}$ converges to the 
square of the overall spectroscopic completeness of the sample 
($f_s^2 \sim 0.78$), as indicated by the dashed line.  The solid lines 
show our model fit to the data, which is used to correct for the small scale 
spectroscopic incompleteness.  (b) The lower panel shows the data 
after these weights have been applied.  The corrected data points provide 
an excellent fit to the overall spectroscopic completeness (solid line), 
confirming that the small scale incompleteness has been successfully removed.
\label{figsmallang}}
\end{figure}

First, we measure the ratio of spectroscopic to 
photometric pairs ($N_{zz}/N_{pp}$) 
as a function of angular separation $\theta$, 
as shown in the upper panel of Figure~\ref{figsmallang}.  
For a fair sample, in which spectroscopic completeness is independent of 
pair separation, one would expect to find $N_{zz}/N_{pp} \sim f_s^2$, 
where $f_s$ is the overall spectroscopic completeness of the survey.  
Instead, we see a sharp drop in $N_{zz}/N_{pp}$ below 
$55\arcsec$, as expected.  We model this incompleteness by fitting a 
function $g(\theta)$ to these data.  We then multiply each  
spectroscopic pair by a weight $w_{\theta} = f_s^2/g(\theta)$.  The resulting 
corrected values of $N_{zz}/N_{pp}$ are plotted in the lower panel 
of Figure~\ref{figsmallang}.  It is clear that this weighting scheme 
is successful in removing the angular dependence of the small scale 
spectroscopic incompleteness, with the corrected 
$N_{zz}/N_{pp} \approx f_s^2$ at all
relevant angular separations, to within the reported errors.
Therefore, by applying $w_{\theta}$ weights to each detected pair, 
we can remove this very significant bias from our measurements.

\subsection{$N_c(M_r)$ for SDSS Pairs} \label{ncmrsdss}
We now have a catalog of host galaxies for which we can measure close
pair statistics, using statistical weights to correct for spectroscopic 
incompleteness.  Weights are combined using the approach outlined by 
\citet{cnoc2mr}.  First,  the 
number of companions for galaxy $i$, summed over all companions $j$, 
is given by 
\begin{equation} \label{eqnnci}
N_{c_i} = \sum_j f^{-1}_{s_j} w_{\theta_{ij}}.
\end{equation}
The statistical weights in this equation are used to 
correct the observed number of companions to the number that would have
been observed in a complete redshift survey.
The mean number of companions per galaxy, weighted by spectroscopic 
completeness, is then given by
\begin{equation} \label{eqnnc}
N_c = \frac{\sum_i f_{s_i} N_{c_i}}{\sum_i f_{s_i}}.
\end{equation}
This weighting scheme places greater importance on galaxies in 
regions of higher
spectroscopic completeness, thereby minimizing statistical uncertainties.

\begin{figure}
\plotone{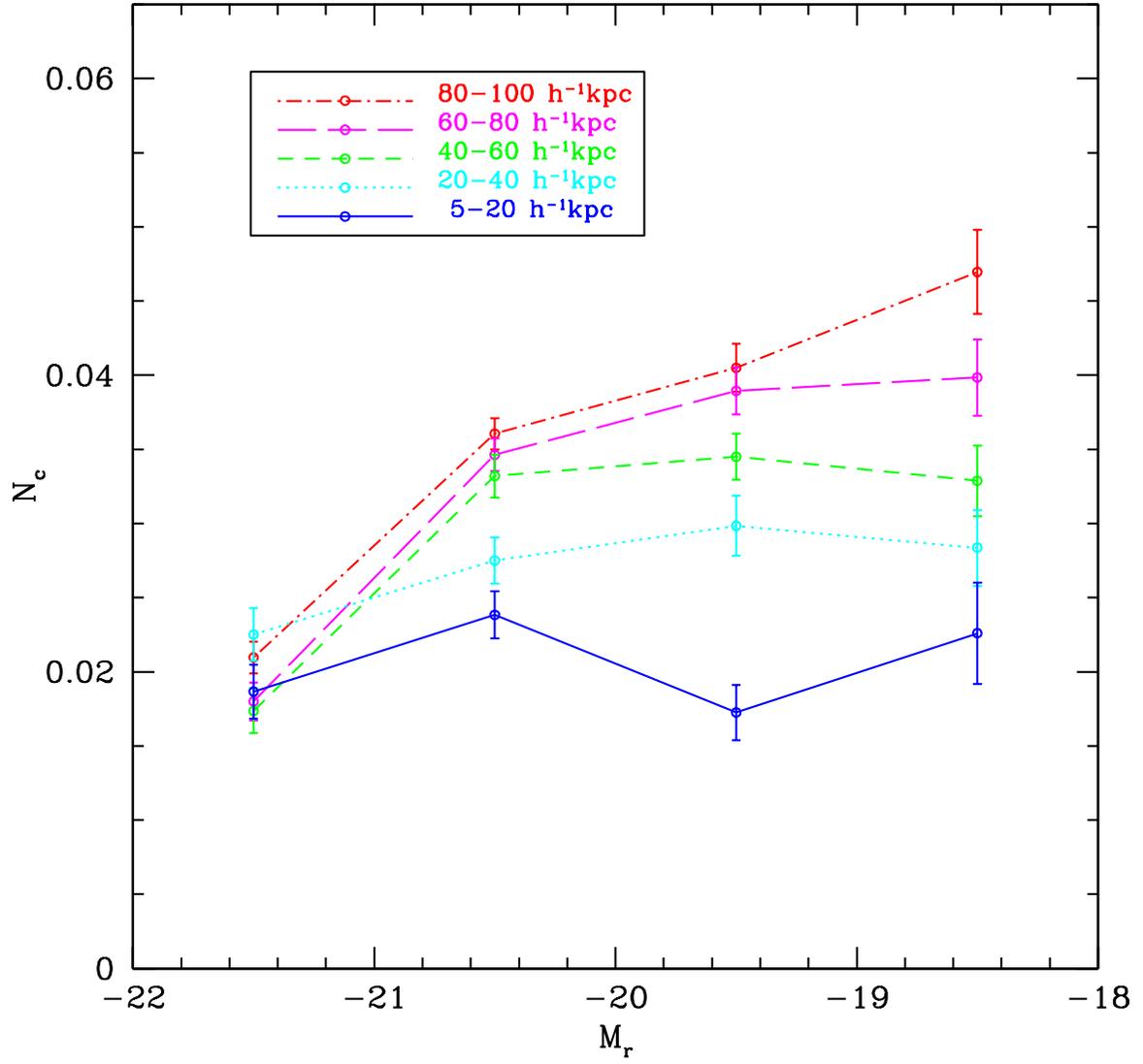}
\caption{$N_c$ is plotted versus absolute magnitude for SDSS, for pairs 
in five ranges of projected separation.  
Over the range $-22 < M_r < -18$, the 
mean $N_c$ for close pairs (5-20~$h^{-1}$ kpc) is $0.021 \pm 0.001$.
The $N_c$ distribution is approximately 
flat for the smallest separation pairs, and becomes 
steeper as pair separation increases.
\label{figncsdss}}
\end{figure}

The resulting measurements of $N_c(M_r)$, calculated using
equations~\ref{eqnnci} and \ref{eqnnc}, 
are indicated by the solid (blue) curve in Figure~\ref{figncsdss}.  
In addition, to facilitate future comparisons with other surveys, 
we provide these results in the first row of Table~\ref{tabsdssnc}.  
We find that
$N_c(M_r)$ is approximately constant 
over the given range in absolute magnitude, with a mean of 
$N_c = 0.021 \pm 0.001$. 
Given that few galaxies have more than one close companion, this implies
that $\sim$ 2\% of galaxies with $-22 < M_r < -18$ have a 
close companion, independent of $M_r$.

Overall, our mean value of $N_c$ agrees quite well with related measurements 
in the literature.  For example, using identical $r_p$ and $\Delta v$ 
criteria, but no limits on luminosity ratio, \citet{ssrs2mr} find 
$N_c(-21 \leq M_B \leq -18) = 0.0226 \pm 0.0052$ at $z = 0.015$, 
while \citet{depropris05} find $N_c=0.0357 \pm 0.0027$ at $z = 0.116$.
Given our additional requirement that companions be within a factor of two 
in luminosity of their host galaxy, both results are broadly consistent with 
the somewhat lower $N_c$ that we find.  
Our mean $N_c$ appears to be several times larger than the
SDSS pair fraction of \citet{kartaltepe07}.  However,  
their close pair criteria are substantially different 
from ours; in 
particular, they require $M_V > -20$ for both members, they do not 
require spectroscopic redshifts for both members of their pairs, and
their sample is derived from the \citet{allam04} sample of merging pairs, 
which have a number of additional criteria imposed (including isolation).
A meaningful comparison with \citet{kartaltepe07} is therefore not feasible.

The flat shape of $N_c(M_r)$ that we find is a surprising result.
Based on the LF alone, one would expect to find 
more companions (of comparable luminosity) 
close to intrinsically faint galaxies, since dwarf galaxies
are much more numerous than giants.  In other words, $N_c$ should 
be proportional to the number density of galaxies \citep{berrier06}.
However, the number of companions 
per galaxy is sensitive to both number density {\it and} clustering strength 
\citep{ssrs2mr,berrier06}.
Clustering strength is known to increase with luminosity 
\citep{norberg01,norberg02,zehavi02,li06}.
This provides a competing effect, which must be 
comparable in size to the density effect if it is to explain the 
flat $N_c(M_r)$ that we find.  However, at separations of 100 $h^{-1}$ kpc, 
the correlation function measurements of \citet{li06} indicate that 
the clustering 
strength of SDSS galaxies is roughly independent of absolute magnitude
for $M_r \gtrsim -21$, and is approximately twice as high in the 
range $-22 < M_r < -21$.  In order to explain the roughly flat $N_c(M_r)$ 
we detect over the range $-22 < M_r < -18$, the luminosity dependence of
clustering must be considerably stronger on the smaller scales relevant
for our close pairs ($\sim$ 10 $h^{-1}$ kpc).
However, we must also rule out any potential luminosity-dependent 
biases in our sample, which could in principle contribute.

It is possible to break the degeneracy between density and clustering 
effects by considering wider separation 
pairs, since clustering strength diminishes with pair separation.  
We therefore compute $N_c(M_r)$ as a function of pair separation out to 
100 $h^{-1}$ kpc (see Figure~\ref{figncsdss} and Table~\ref{tabsdssnc}).  
As expected, $N_c(M_r)$ steepens towards the faint end as pair separation 
increases.  
This implies that $N_c(M_r)$ scales 
with number density on large scales, but flattens out on small scales as 
a result of the added effects of luminosity-dependent clustering.
We conclude that the roughly flat $N_c(M_r)$ that we observe for close 
pairs is seen only on the small scales relevant for galaxy interactions and 
mergers, and is unlikely to be due to any overall luminosity-dependent 
bias within our sample.  

\section{Millennium Close Pair Statistics}\label{ncmill}

Given the somewhat surprising trends seen in our SDSS pair statistics, 
and to facilitate comparison with theoretical models of galaxy formation 
and evolution, we now apply our techniques to a sample 
drawn from the semi-analytic galaxy catalogs of \citet{croton06}, which 
were created using the output of the Millennium Run simulation 
\citep{springel05}.  
These catalogs have been shown to reproduce 
the overall properties of galaxies in the local universe, including 
the luminosity function, the two point correlation function, and the pairwise 
velocity dispersion \citep{croton06,li07}.
The simulations provide the additional benefit 
of three dimensional positions and velocities, which we will use to 
assess and remove projection effects.  While the reader is referred to 
\citet{croton06} and references therein for a detailed description of these 
catalogs, we note that merging is treated by following dark matter subhalos 
down to a mass limit of $1.7 \times 10^{10} h^{-1} \msun$, and then using 
the dynamical friction formula of \citet{binney87} to estimate the 
remaining time until a merger takes place.

For this analysis, we use the \citet{croton06} $ugriz$ catalog, which is 
complete for galaxies more luminous than $M_r = -16.6$.  
The stated resolution is 5 $h^{-1}$kpc, which conveniently coincides 
with our minimum pair separation criterion.  We confirm, however, that 
the two-point galaxy correlation function has the expected power law 
form to below this level, so there should not be any unexpected effects 
near the resolution limit.

\subsection{Mock Redshift Catalogs}\label{mock}

The \citet{croton06} catalogs contains three dimensional positions and 
velocities for approximately 9 million galaxies at redshift zero, 
within a cube which is 500 $h^{-1}$~Mpc on a side.  
In order to make a direct comparison between Millennium and SDSS, 
it is necessary to transform this real-space catalog into a mock redshift 
catalog.  To this end, we begin by
placing the observer at the origin of the Millennium cube,
and then computing the corresponding right ascension, declination, 
redshift, and apparent magnitude of each galaxy in the sample.  In order to 
match our observed SDSS sample, we then apply the same flux limits
($14.5 < m_r < 17.5$).  In addition, we impose a minimum redshift of 
0.022, to ensure that we sample only galaxies which are more luminous than 
$M_r = -16.6$ (the completeness limit of the Croton catalog).  
Finally, we impose a maximum redshift of 0.17, which ensures that we 
do not probe distances greater than 500 $h^{-1}$~Mpc (the size of the 
simulation cube).
The resulting catalog contains roughly 300,000 
galaxies, or about one thirtieth of the full \citet{croton06} sample. 

There are two disadvantages to this simple approach.  First, the vast majority 
of the galaxies in the simulation are discarded as a result of the 
imposed flux limits, despite the fact that they are sufficiently 
luminous to be of interest.  Secondly, in order to assess projection effects, 
it would be useful to view galaxy associations from a variety of vantage 
points.  

Fortunately, the periodic boundary conditions imposed on the 
simulations provide a way forward.  Following part of the 
``random tiling'' technique outlined by \citet{blaizot05}, 
we generate a suite of mock redshift catalogs.  
Specifically, for each mock catalog, we begin by applying to the Millennium
cube a 
translation in each of the three spatial directions, 
with the size of each translation being a random fraction of the box size 
(500 $h^{-1}$~Mpc).  For any resulting coordinate value greater than 
the box size, we subtract the box size, thereby ``wrapping around'' 
the cube.  Finally, we rotate the cube around each of the three spatial 
axes by a random (integer) multiple of $\pi/2$.  
Using this approach, we create a suite of 30 mock redshift catalogs, 
sampling a total of about 9 million galaxies (comparable to the number 
of galaxies in the original data cube).

\subsection{Redshift-space Pair Statistics}\label{ncmrmill}

We then proceed to compute 
pair statistics on all of the mock redshift catalogs, selecting
host and companion galaxies in the same manner outlined 
in Sections~\ref{methodology} and \ref{hostcomp}.
One key difference in 
technique must be noted, however: given that the Millennium data 
is a redshift zero realization, we compute co-moving rather than 
physical projected separations in order to 
recover the correct $z=0$ physical separations.

\begin{figure}
\plotone{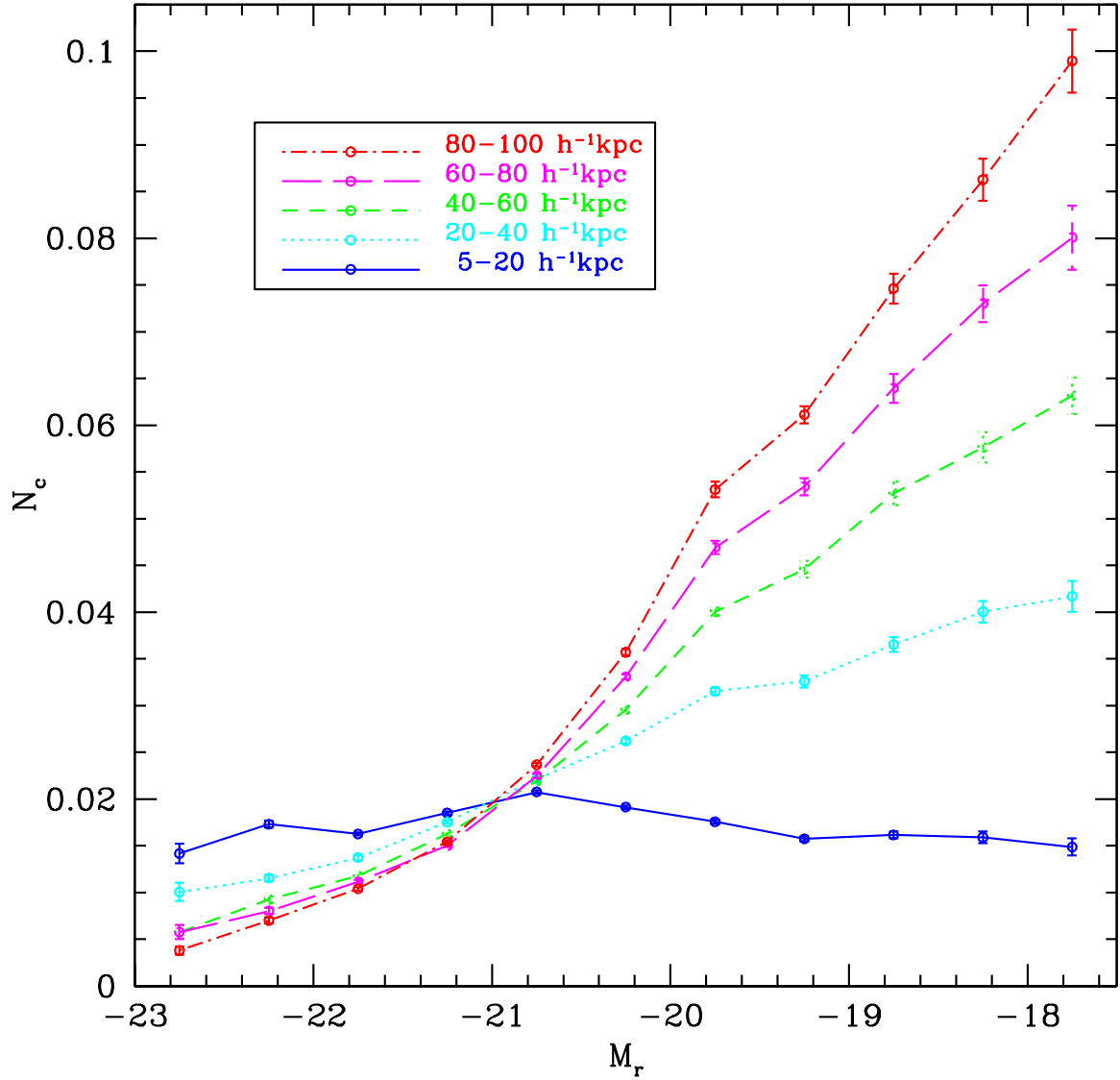}
\caption{$N_c$ is plotted versus absolute magnitude for the Millennium 
simulation, for pairs in five ranges of projected separation.  
As with SDSS, the $N_c$ distribution is found to be approximately flat for the 
smallest pair separations.  
Over the range $-22 < M_r < -18$, the 
mean $N_c$ for close pairs is $0.0183 \pm 0.0001$, 
which is about 15\% lower than found for SDSS.
The Millennium $N_c$ distribution steepens towards the faint end
as pair separation increases.  This is similar to what was seen for SDSS, 
but is considerably steeper.
\label{figncmill}}
\end{figure}

Our results are presented in Figure~\ref{figncmill} and in the first two 
columns of 
Table~\ref{tabmill}.  Overall, the trends
seen are broadly similar to those for SDSS shown in Figure~\ref{figncsdss}.
The distribution is again relatively flat for the smallest separation 
pairs, and increases towards the faint end as pair separation increases.
This implies that the projected 
two point correlation becomes steeper on small scales 
as luminosity increases, which is consistent with the increase in 
steepness with stellar mass found for 
Millennium galaxies by \citet{kitzbichler08}.
For close pairs ($5 < r_p < 20~h^{-1}$ kpc), 
$N_c(M_r)$ peaks at a value of about 0.02 
at $M_r \sim -20.75$.  Given that $M^*$ is approximately $-20.6$ in the 
$r$ filter \citep{blanton03a}, 
this implies that $M^*$ galaxies are the most likely 
to have close companions.  For close pairs over the range $-22 < M_r < -18$, 
the mean $N_c$ is $0.0183 \pm 0.0001$, 
which is about 15\% smaller than we found for SDSS.  
The most significant 
difference between SDSS and Millennium is seen for the larger separation 
pairs, in that the Millennium $N_c(M_r)$ is considerably steeper.  
We discuss a possible explanation for this in Section~\ref{ncv}.
However, given that we are primarily interested in the smallest separation 
pairs, the general agreement seen between SDSS and Millennium in this regime
justifies further comparison between these samples.

\subsection{Projection Effects}\label{f3d}

One of the most challenging aspects of close pair studies is the 
contamination of pair samples due to projection effects.  
For any observed galaxy pair, even when spectroscopic 
redshifts are available for
both galaxies, one cannot be certain that the galaxies are close 
enough to merge.  In addition, without measurements of transverse velocities, 
it is also possible that the relative velocities of the member galaxies
are too high for coalescence to occur.  For pairs with 
$5 < r_p < 20~h^{-1}$ kpc and $\Delta v < 500$ km s$^{-1}$, 
\citet{ssrs2mr} estimated this contamination to be on the order of 50\%.
One must apply such a statistical correction for projection effects in order 
to relate measured pair statistics to their real-space (three-dimensional) 
equivalents.

The three dimensional information available in the \citet{croton06} 
catalogs enables us to instead measure the level of contamination directly.
For every pair of galaxies that we observe in our mock redshift catalogs, 
we can measure the three dimensional physical separation, along with 
the three dimensional relative velocity.  If these quantities are 
less than $20~h^{-1}$ kpc and 500 km s$^{-1}$ respectively, it is likely 
that a merger is imminent \citep{ssrs2mr}.  
In addition, every pair satisfying these 
three dimensional criteria will also satisfy the redshift space 
criteria (i.e., in terms of $r_p$ and $\Delta v$) when viewed from 
{\it any} vantage point.

For our Millennium mock catalogs, we measure the fraction of close 
companions which satisfy the three dimensional criteria; following 
\citet{ssrs2mr}, we refer to this fraction as $f_{3D}$.  
Over the range $-22 < M_r < -18$, we find an overall $f_{3D}$ of 
$47.4\% \pm 0.3\%$.  This compares well with the rough estimate of 
$f_{3D} \sim 50\%$ of \citet{ssrs2mr}, although 
their estimate applies to different absolute 
magnitude criteria ($-21 < M_B < -18$ for both hosts and companions).

\begin{figure}
\plotone{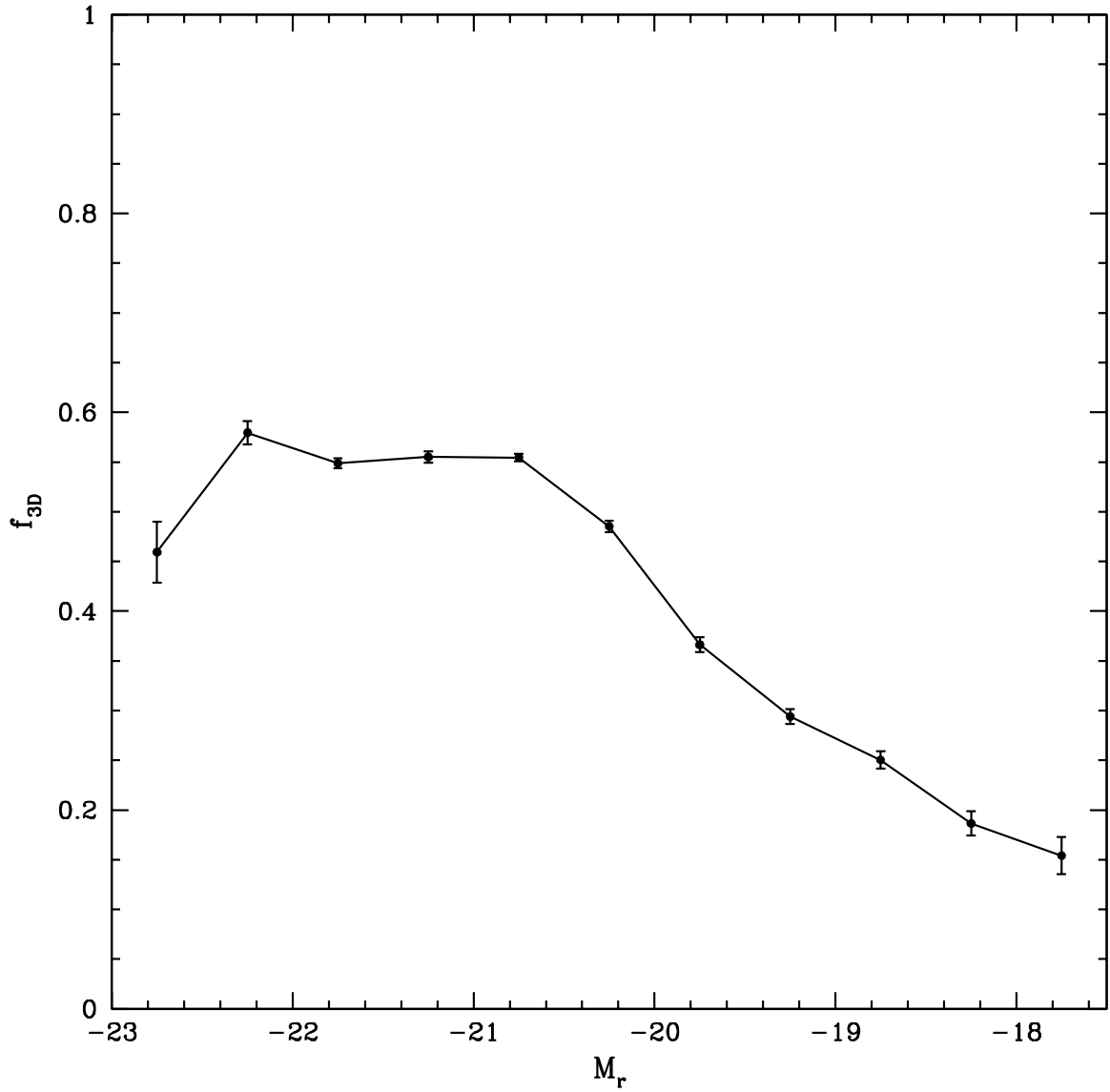}
\caption{Using the Millennium simulation, 
the fraction of companions which are close in three dimensions 
($f_{3d}$)
is plotted versus absolute magnitude for close pairs ($5-20~ h^{-1}$ kpc).
Projection effects are seen to increase towards fainter absolute magnitudes.
\label{figf3d}}
\end{figure}

In addition to measuring the overall $f_{3D}$, it is also useful to 
probe the dependence of this quantity on absolute magnitude.  If 
it varies significantly with $M_r$, this will have implications for
any merger rate estimates we wish to glean from our $N_c(M_r)$ measurements.
The results are given in Figure~\ref{figf3d}, and in the third column 
of Table~\ref{tabmill}.  
Interestingly, $f_{3D}$ is seen to vary by a factor of about three
over the range in $M_r$ probed, decreasing from $\sim 56\%$ 
at $M_r \lesssim M^*$ to about 17\% at $M_r \sim -18$.  
This implies that most of the lowest luminosity close pairs in our Millennium 
mock catalogs are not candidates for imminent mergers, 
while at least half of the close
pairs with $M_r \lesssim M^*$ are likely to undergo imminent mergers. 

\section{Close Pair Statistics in Real Space}\label{realspace}

\subsection{Millennium}\label{nctruemill}
In the preceding section, we reported our measurement of $f_{3D}(M_r)$ from 
the Millennium simulation.  Multiplying this function by the 
redshift space close pair statistics reported in \S~\ref{ncmrmill}, we arrive
at {\it real space} pair statistics, in which projection effects 
have been removed.  In Figure~\ref{fignc20}, we present these results 
for the Millennium simulation, along with the redshift-space results 
given earlier.  We also provide tabulated values in column 4 of 
Table~\ref{tabmill}.
In real space, we find that $N_c(M_r)$ peaks at $M_r \sim M^*$, 
at a value of $\sim$~0.011, rather than $\sim$~0.02 in redshift space.
Also, rather than being relatively flat (as it is in redshift space), 
we now see a strong decline towards fainter luminosities, as 
a result of the associated decline in $f_{3D}$.  The real space 
$N_c$ drops to about 0.002 at the faintest luminosities probed, which is 
much lower than in redshift space.  These results imply that
$L^*$ galaxies are much more likely to have close companions than 
$0.1L^*$ galaxies.  This striking difference is not apparent in 
redshift space, due to the presence of projection effects.
We also see a hint that the real space $N_c$ declines at 
the bright end ($M_r \sim -23$), though this is seen only in the 
most luminous bin in the simulations. 

\subsection{SDSS}

The strong contribution from projection effects seen in the 
Millennium simulation has important implications for the SDSS redshift space
pair statistics presented in \S~\ref{ncmrsdss}.  In particular, the
relatively flat $N_c(M_r)$ observed for SDSS close pairs almost certainly 
does not hold true in real space.  Without three dimensional positions 
and velocities for SDSS galaxies, we are unable to directly remove 
projection effects.  However, 
given the general agreement between 
the SDSS and Millennium redshift space pair statistics (see \S~\ref{ncmrmill}), 
it seems reasonable to apply our knowledge of Millennium projection effects 
to our SDSS pair statistics.  Multiplying the 
SDSS redshift space $N_c(M_r)$ 
by the Millennium $f_{3D}(M_r)$  
yields our 
best estimate of SDSS close pair statistics in real space.  
These results are given in Figure~\ref{fignc20}, as well as in 
Table~\ref{tabsdss}.  We find that the SDSS real space $N_c(M_r)$ 
is approximately 0.011 for $-22 < M_r < -20$, and then drops to 
about 0.0055 for $-20 < M_r < -18$.  
Given the larger error bars in our SDSS measurements, 
we cannot state with any certainty where the real-space $N_c(M_r)$ 
peaks; however, our results are consistent with a peak at $M_r < -20$, 
as seen in the Millennium simulations (\S~\ref{nctruemill}).

\subsection{Comparison with Other Studies}\label{nccompare}

Most observed samples of close pairs have been too small to permit anything 
meaningful to be learned about the luminosity (or mass) dependence of 
close pair statistics.  
The most notable exception is
\citet{xu04}, who measure the pair fraction as a function of
$K_s$-band absolute magnitude for galaxies in the 2MASS survey.
While they generally probe higher luminosities ($-24.5 < M_K < -22.5$) 
than we do, their two faintest bins (centered on $M_k = -22$) correspond
roughly to our brightest SDSS bin ($M_r = -21.5$); in this regime, they find
a pair fraction of about $0.011 \pm 0.005$, which is consistent
(within 1$\sigma$) of our $N_c \sim 0.019 \pm 0.002$ (Table~\ref{tabsdss}), 
despite some differences between their pair criteria and ours.
However, \citet{xu04} find that the pair fraction increases towards 
higher luminosities (albeit with large error bars), whereas we see 
a hinted of a decrease with luminosity in this regime with Millennium.
On the other hand, our observed trend is consistent with the 
semi-analytic results 
of \citet{khochfar01}, who find that the $z=0$ merger fraction decreases 
as mass increases, over a range in masses which corresponds to 
$M_r \lesssim M^*$.  

We also compare our 
results with those of \citet{berrier06}, who measure $N_c$ 
using a combination of N-body simulations and semi-analytic models.  
They find that $N_c$ increases steadily as the cumulative co-moving
number density increases.  This is equivalent to an increase in $N_c$ 
towards fainter limiting $M_r$, 
over a range corresponding to $M_r \gtrsim M_r^*$
(using the SDSS LF measurements of \citet{blanton03a}).  We instead see a 
steady decrease in $N_c$ towards fainter $M_r$ over this range.  
However, \citet{berrier06} 
include all companions brighter than the given number density 
(or equivalently, $M_r$).  With this approach, the cumulative number of 
close companions for a given galaxy can only go up as $M_r$ 
becomes fainter.  It is therefore not surprising that their 
measurements of $N_c$ rise towards fainter $M_r$.  Given that 
we require companions to be within a factor of two in luminosity 
of their host galaxy, there is no reason to expect agreement between 
our measurements of $N_c$ and those of \citet{berrier06}.  We note, however, 
that our definition of $N_c$ is likely to be a better tracer of 
the {\it major} merger rate.

\section{From Close Pair Statistics to the Merger Rate}\label{mrate}

\subsection{The Number Density of Close Companions}\label{ncv}

We have measured the number of companions per galaxy ($N_c$), which 
tells us which galaxies are most likely to have close companions.
The real-space distributions seen in Figure~\ref{fignc20} indicate 
that low luminosity galaxies are the least likely to have close 
companions, and presumably the least likely to undergo 
imminent mergers.  However, low luminosity galaxies are much more
common than luminous galaxies, and therefore it is still possible 
that most major mergers may occur between low luminosity galaxies.

We investigate this question by measuring the number of close companions
per unit co-moving volume, hereafter $n_c$.  In order to compute 
this quantity, we multiply the number of close companions per 
galaxy ($N_c(M_r)$ by the co-moving number density of galaxies ($n(M_r)$).
For SDSS, we estimate $n(M_r)$ using the LF measurements
of \citet{blanton03a}, converting from the $^{0.1}r$ filter to the $r$ filter
using their recommended $M_r = M_{0.1_r} - 0.16$.  For Millennium, 
we measure the galaxy number density directly from the \citet{croton06} 
galaxy catalog.  Like \citet{li07}, we find that the \citet{blanton03a} 
LF matches the \citet{croton06} catalog quite well overall, 
though the simulations overpredict the number of galaxies at fainter 
luminosities (the excess is $\sim 50\%$ at $M_r = -18$).  We note that 
this excess helps to explain why $N_c(M_r)$ for wide separation 
pairs is steeper for Millennium than for SDSS (see \S~\ref{ncmrmill}).

The resulting measurements of $n_c(M_r)$ 
are given in Figure~\ref{fignc} and in Tables~\ref{tabsdss} 
and \ref{tabmill}.  Excellent agreement is seen between 
SDSS and Millennium.  When summed over the range $-22 < M_r < -18$, 
the SDSS $n_c = (2.10 \pm 0.13) \times 10^{-4} h^3 {\rm Mpc}^{-3}$, which is 
consistent (within one sigma) with the Millennium 
$n_c = (1.98 \pm 0.02) \times 10^{-4} h^3 {\rm Mpc}^{-3}$.
In addition, the shapes of the distributions are very similar, 
with $n_c \sim 6 \times 10^{-5} h^3 {\rm Mpc}^{-3} {\rm mag}^{-1}$ 
for $M_r > -21$, and a sharp drop in $n_c$ at higher luminosities.  
For $M_r > -21$, it therefore appears that the decrease in the 
real-space $N_c(M_r)$ towards fainter absolute magnitudes is balanced by 
a corresponding increase in host galaxy number density.  On the other hand, 
the relative scarcity of luminous host galaxies leads to the rapid decline in 
$n_c$ towards bright absolute magnitudes.  From Figure~\ref{fignc}, 
we conclude that at least 90$\%$ of all 
major mergers occur between galaxies which are fainter than $M_r = -21$.
We note that, while there are no other directly comparable measurements
in the literature, the shape of our observed $n_c(M_r)$ is qualitatively 
similar to the shape of the merging galaxy mass functions of 
\citet{bundy05} and \citet{hopkins06}.

\subsection{The Galaxy Merger Rate}

\begin{figure}
\plotone{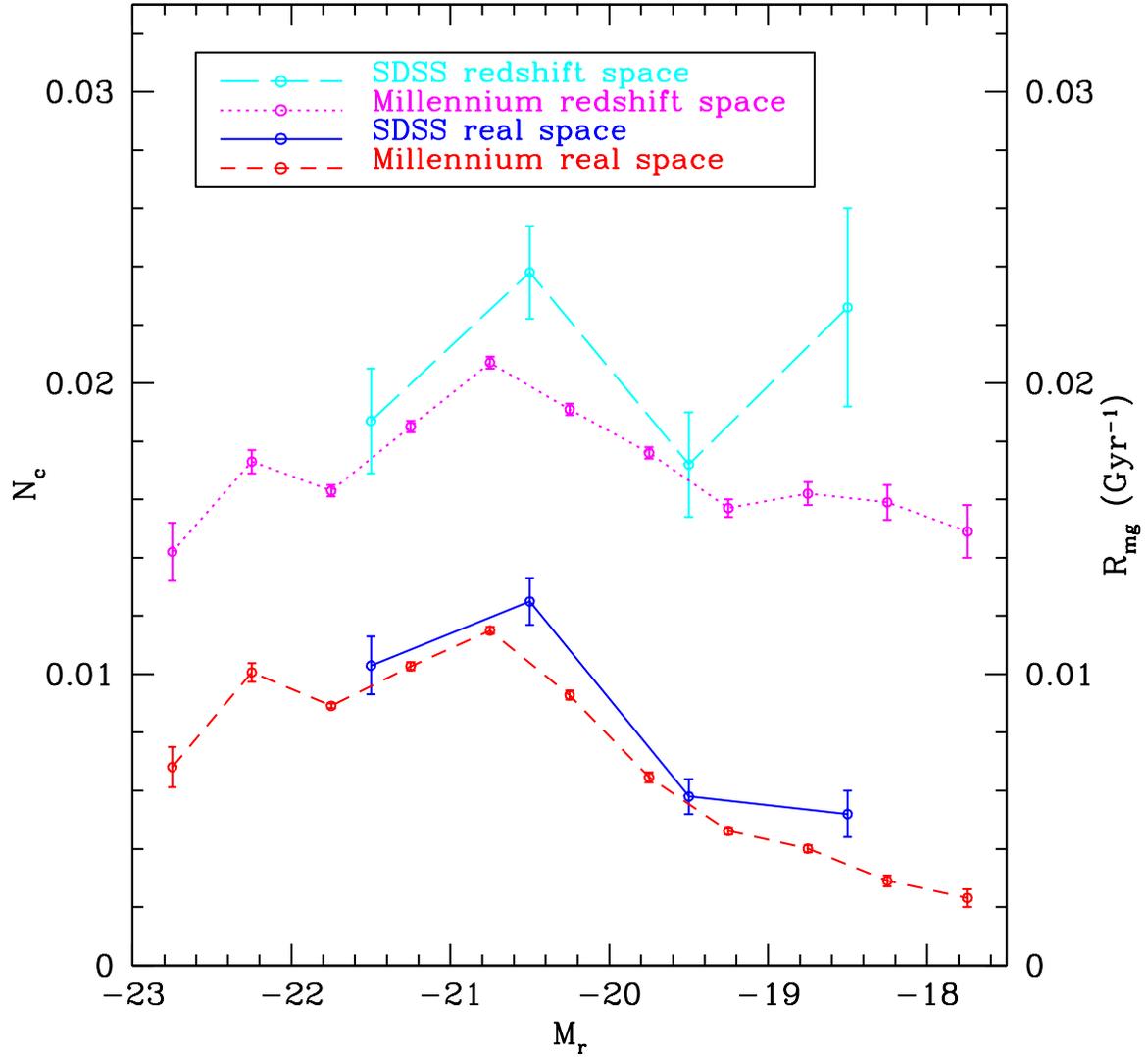}
\caption{$N_c(M_r)$ is plotted 
in both redshift space and real space, for close pairs ($5-20~ h^{-1}$ kpc).
For both SDSS and Millennium, conversion from redshift space to real space 
pair statistics (ie., the removal of projection effects) was carried out
by multiplying $N_c(M_r)$ by the Millennium $f_{3D}(M_r)$.
The right hand axis displays the corresponding 
scale for the galaxy merger rate ($\rmg$), and is relevant only for 
measurements in real space.
\label{fignc20}}
\end{figure}

\begin{figure}
\plotone{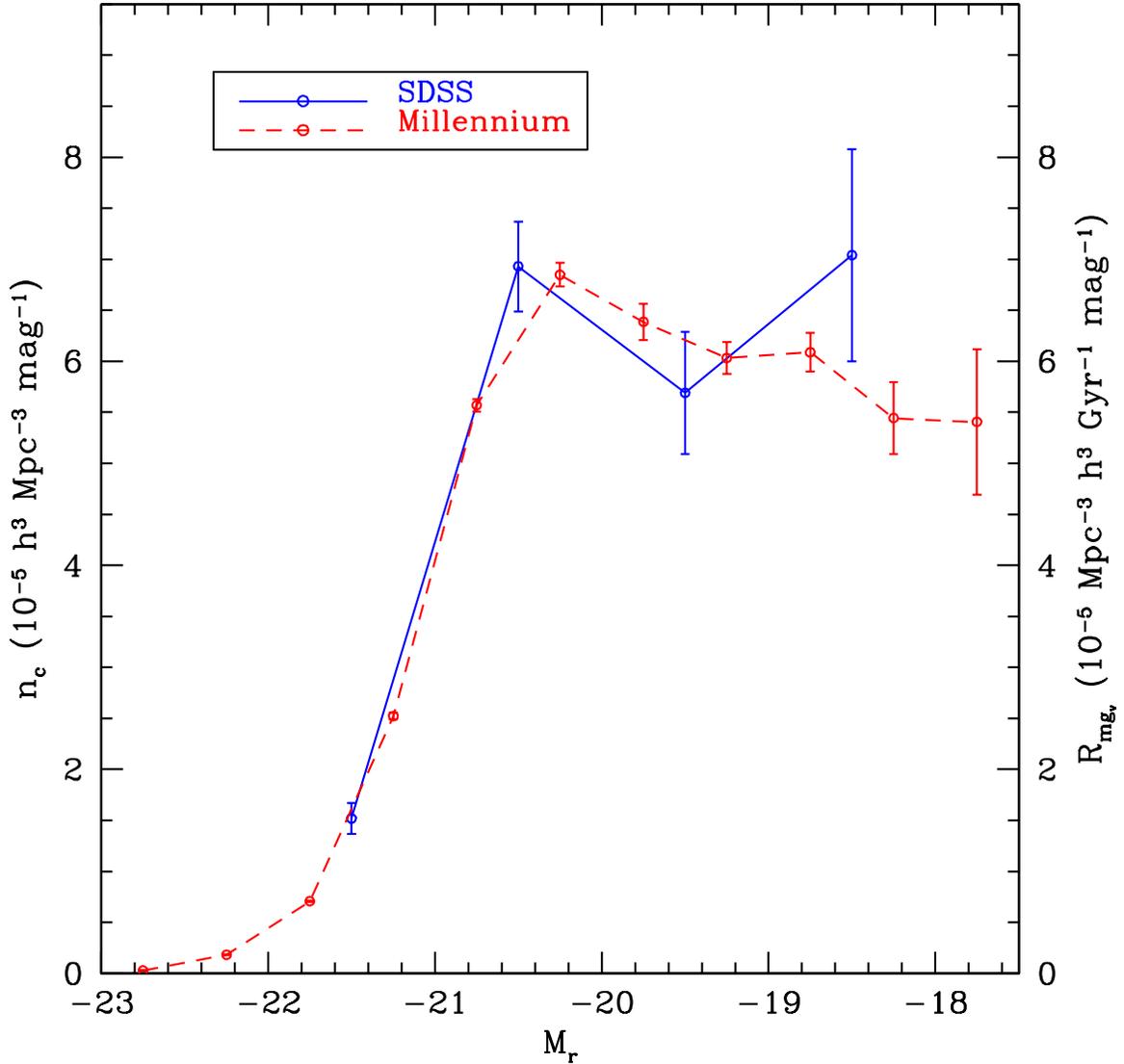}
\caption{The number of close companions per unit co-moving volume ($n_c$) 
is plotted versus $M_r$ for SDSS and for Millennium.
In both cases, $n_c$ is computed by multiplying the real-space $N_c$
by the number density of galaxies.  
Excellent agreement is 
seen between the two samples.  We conclude that most mergers occur between 
galaxies with $M_r > -21$.  The right hand axis displays the 
corresponding volume merger rate.
\label{fignc}}
\end{figure}

We have reported estimates of the number of companions per galaxy ($N_c$) 
and per unit volume ($n_c$).  With additional assumptions, both 
can be converted to merger rates.  Following the formalism 
of \citet{ssrs2mr}, we will refer to these as the galaxy merger rate 
(hereafter $\rmg$) and the volume merger rate ($\rmgv$) respectively.
Assuming that two real-space close companions (ie., one galaxy pair) 
lead to one merger, 
and that the average timescale for such mergers is $T_{\rm mg}$, 
it follows that 
$\rmg = 0.5N_c/T_{\rm mg}$ and $\rmgv = 0.5n_c/T_{\rm mg}$, 
where both $N_c$ and $n_c$
refer to real-space measurements (Figures~\ref{fignc20} and \ref{fignc}).

To proceed further requires an estimate of $T_{\rm mg}$.   
We begin by taking the simplest approach, which is to assume that 
$T_{\rm mg}$ is equal to the dynamical friction timescale of 
a typical pair in the sample (after projection effects have been removed).
We adopt the estimate of 0.5 Gyr given by \citet{ssrs2mr}, since
our pairs are chosen with similar criteria.  This timescale estimate is 
comparable to or somewhat larger than several more recent 
estimates given in the literature \citep{vandokkum05,bell06a,depropris07}.
While the merger timescale may depend on mass (and hence 
absolute magnitude), N-body simulations appear to indicate that 
mass ratio is the most significant factor \citep{jiang08}, and this 
should be independent of $M_r$ for the pair criteria we employ.

The resulting merger rates are plotted using the right hand 
axes in Figures~\ref{fignc20} and \ref{fignc}.  The galaxy merger 
rate is found to peak at about 0.01 mergers per galaxy per Gyr, 
while the volume merger rate plateaus at about 
$6 \times 10^{-5} h^3 {\rm Mpc}^{-3} {\rm Gyr}^{-1} {\rm mag}^{-1}$. 
As the conversion 
from pair statistics ($N_c$ and $n_c$) to merger rates ($\rmg$ and $\rmgv$) 
is independent 
of $M_r$, the trends with $M_r$ described in Sections~\ref{realspace} and 
\ref{ncv} apply here as well.  We again conclude that while galaxies 
brighter than $L^*$ have the highest likelihood of being involved in 
major mergers, most major mergers take place between 
galaxies fainter than $L^*$.  

How do these merger rate estimates compare with others in the literature?
\citet{depropris07} estimated the merger rate using two methods: 
close galaxy pairs 
(where both galaxies were required to have $-21 < M_B < -18$) 
and high asymmetries 
(as an indicator of ongoing mergers).  They found a merger fraction 
close to $2\%$ for both methods, which corresponds to 
$\rmg \sim 0.02~{\rm Gyr}^{-1}$ using our $T_{\rm mg} = 0.5$ Gyr.  
This is approximately twice as high as our result.
\citet{depropris07} also 
report a volume merger rate of 
$(5.2 \pm 1.0) \times 10^{-4} h^3 {\rm Mpc}^{-3} {\rm Gyr}^{-1}$, 
whereas we find 
$\rmgv = (1.4 \pm 0.1) \times 10^{-4} h^3 {\rm Mpc}^{-3} {\rm Gyr}^{-1}$ when 
summing over a comparable range in absolute magnitude ($-22 < M_r < -19$).
After recomputing their result with our $T_{\rm mg} = 0.5$ Gyr (instead of
their 0.3 Gyr), we again find that their merger rate is roughly double ours.
The most likely explanation for this difference is that both the pairs 
and asymmetry methods of \citet{depropris07} include both major and 
minor mergers, whereas our measurements apply to major merger candidates 
only. 

\citet{maller06} measure the galaxy merger rate as a function of mass, 
mass ratio, and redshift, using a smoothed particle hydrodynamics (SPH) 
cosmological simulation.  For mergers above a 1:2 mass ratio (comparable to 
our 1:2 luminosity ratio), they find the galaxy merger rate at $z=0.3$ 
to be $\rmg = 0.054$ Gyr$^{-1}$ for high mass galaxies.  This range 
in mass corresponds approximately to $-18.8 < M_r < -17.7$ (derived from 
Table 1 of Maller et al. (2006)\nocite{maller06}). 
While they do not report precise values of $\rmg$ at lower redshifts, 
they do find that $\rmg$ declines quickly towards $z=0$, yielding 
$\rmg \lesssim 0.02$ Gyr$^{-1}$ at $z < 0.1$.  
This value is consistent with the 
$\rmg \sim 0.005$ Gyr$^{-1}$ we find at $z \sim 0.03$ over this 
range in $M_r$, particularly given that no error bars on the \citet{maller06} 
estimate are available.
Interestingly, they find that the merger rate is lower 
for lower mass galaxies, in agreement with the trends we find 
in Figure~\ref{fignc20} (particularly for Millennium).
\citet{maller06} also measure a volume merger rate of about 
$1 \times 10^{-4} h^3 {\rm Mpc}^{-3} {\rm Gyr}^{-1}$ (with large error bars) 
for high mass galaxies at $z=0.1$; this is consistent with the 
$\rmgv \sim 6 \times 10^{-5} h^3 {\rm Mpc}^{-3} {\rm Gyr}^{-1}$ 
we find in Figure~\ref{fignc}.  In general, \citet{maller06} find that 
$\rmgv$ declines towards lower masses (though this effect is significant 
only at $z \gtrsim 0.4$), implying that one might expect to see $\rmgv$ 
decline at fainter absolute magnitudes than probed by our SDSS and 
Millennium catalogs.

\citet{masjedi06} measured the volume merger rate for pairs of luminous 
red galaxies (LRG's) in the SDSS, and found 
$\rmgv \lesssim 0.6 \times 10^{4} {\rm Gyr}^{-1} {\rm Gpc}^{-3}$.  
Both members of their LRG's pairs 
were required to have absolute magnitudes of $-23.2 < M_g < -21.2$, 
which is different in nature 
from our major merger requirement of $|\Delta M_r| < 0.753$.
While our SDSS pairs are less luminous than the \citet{masjedi06} LRG's, 
our Millennium volume merger rate estimates in Figure~\ref{fignc} 
yield a similar result 
($\rmgv \lesssim 1 \times 10^{4}~{\rm Gyr}^{-1} h^3 {\rm Mpc}^{-3}$)
in the LRG luminosity range.
Finally, we note that while \citet{masjedi08} probe the luminosity dependence
of close companions of LRG's in the SDSS, their (LRG) host 
galaxies are more luminous than the host galaxies we probe here, and 
most of their close companions are not luminous enough to qualify as 
major merger candidates; therefore, there is essentially no overlap between our 
studies.  However, their Figure 3 is consistent with a major merger rate of
$\rmg \sim 0.001 {\rm Gyr}^{-1}$ for typical LRG's,
which is in qualitative agreement with 
the hinted drop of $N_c(M_r)$ towards high luminosities which we describe 
in \S~\ref{nccompare}.

We caution the reader that our estimates of the merger rate 
depend on two quantities: the close pair statistics we have {\em measured}
($N_c$ or $n_c$), and the merger timescale ($T_{\rm mg}$) that we have 
estimated.  The latter is considerably less certain, and different choices 
of the merger timescale and its dependence on mass (or luminosity) may 
significantly change our estimated merger rates.  
The most useful 
check on our results comes from \citet{kitzbichler08}, who use the Millennium 
simulation to devise fitting formulas for estimating the merger timescale for 
pairs within a given projected physical separation (hereafter $T_{\rm merge}$).  
While none of 
their fitting formulas are directly applicable to our sample, the closest 
match comes from using their equation 9 with $r_p < 30~h^{-1}$ kpc, 
$v_p < 300$ km s$^{-1}$, and $z=0$.  
We approximate the limiting stellar mass as the 
median \citet{croton06} stellar mass of galaxies corresponding to the 
faintest allowable companions for a given host galaxy.  We note also that 
the \citet{kitzbichler08} fitting formulas apply to pairs in which  
galaxy stellar masses differ by at most a factor of 4, whereas our sample 
consists of pairs in which galaxy luminosities differ by a most a factor of 2.

These choices yield $T_{\rm merge}$ estimates which increase monotonically 
from about 1 Gyr at $M_r \sim -23$ to 4.1 Gyr at $M_r \sim -17.5$ (the range 
covered by our Millennium sample).  At face value, these results may 
appear to be
at odds with the fixed merging timescale of 0.5 Gyr that we adopted earlier 
in this section.
However, there is an important difference: the \citet{kitzbichler08} 
timescale is relevant for {\it projected} pairs, while our timescale is 
for real space pairs.  To make a direct comparison, our timescale of 0.5 Gyr 
needs to be divided by $f_{\rm 3D}(M_r)$; this leads to values of $T_{\rm merge}$ 
ranging from 0.8 Gyr to 3.2 Gyr over the same range in $M_r$, 
which compare quite favorably with the 
preceding calculations.
Therefore, despite significant mismatches between 
our sample and the \citet{kitzbichler08} fitting formula criteria, 
both approaches lead to a similar luminosity dependence 
of the merging 
timescales for projected pairs, and the overall timescales are 
comparable.  Moreover, our estimates of $f_{\rm 3D}(M_r)$ 
(see Figure~\ref{figf3d}) provide additional insight into the 
nature of the increase in $T_{\rm merge}$ with decreasing luminosity; 
namely, that this trend may be a consequence of most lower luminosity pairs 
having separations which are too large for merging to take place, 
rather than such systems simply undergoing a slower merging process.
Nevertheless, it is clear that more work is needed in order to 
more accurately model the merging timescales of observed pair samples.

\subsection{The Remnant Fraction}

Given the above measurements of the merger rate, it is in principle 
possible to assess the merging history of low redshift galaxies.  
Following \citet{ssrs2mr}, we estimate the 
fraction of galaxies which have undergone major mergers since $z = 1$.
This quantity, which is referred to as the remnant 
fraction ($f_{\rm rem}$), is given by
\begin{equation} \label{eqnfrem}
f_{\rm rem}(M_r) = 1 - \prod_{j=1}^N{1- N_c(M_r,z_j) \over 1 - 0.5 N_c(M_r,z_j)},
\end{equation}
where $N_c(M_r)$ is measured in real space (e.g., \S~\ref{realspace}), 
$z_j$ corresponds to a lookback time of $jT_{\rm mg}$, and $N$ refers to 
the number of merger timescales over the range $0 < z < 1$ 
($N=15$ for our chosen cosmology).
In principle, one needs to measure $N_c(M_r)$ at $0 < z < 1$ in order
to accurately compute $f_{\rm rem}(M_r)$.
As we have only our low redshift ($z < 0.1$) measurements of $N_c(M_r)$ to 
work with, we make the simplifying assumption that $N_c(M_r)$ does 
not evolve with redshift.  If instead $N_c(M_r)$ rises with redshift, 
as indicated by numerous studies (see \S~\ref{intro}), 
the resulting remnant fractions will be higher \citep{cnoc2mr}.   
And of course, if the shape of $N_c(M_r)$ evolves strongly, 
the shape of our estimated $f_{\rm rem}(M_r)$ will also be in error.

\begin{figure}
\plotone{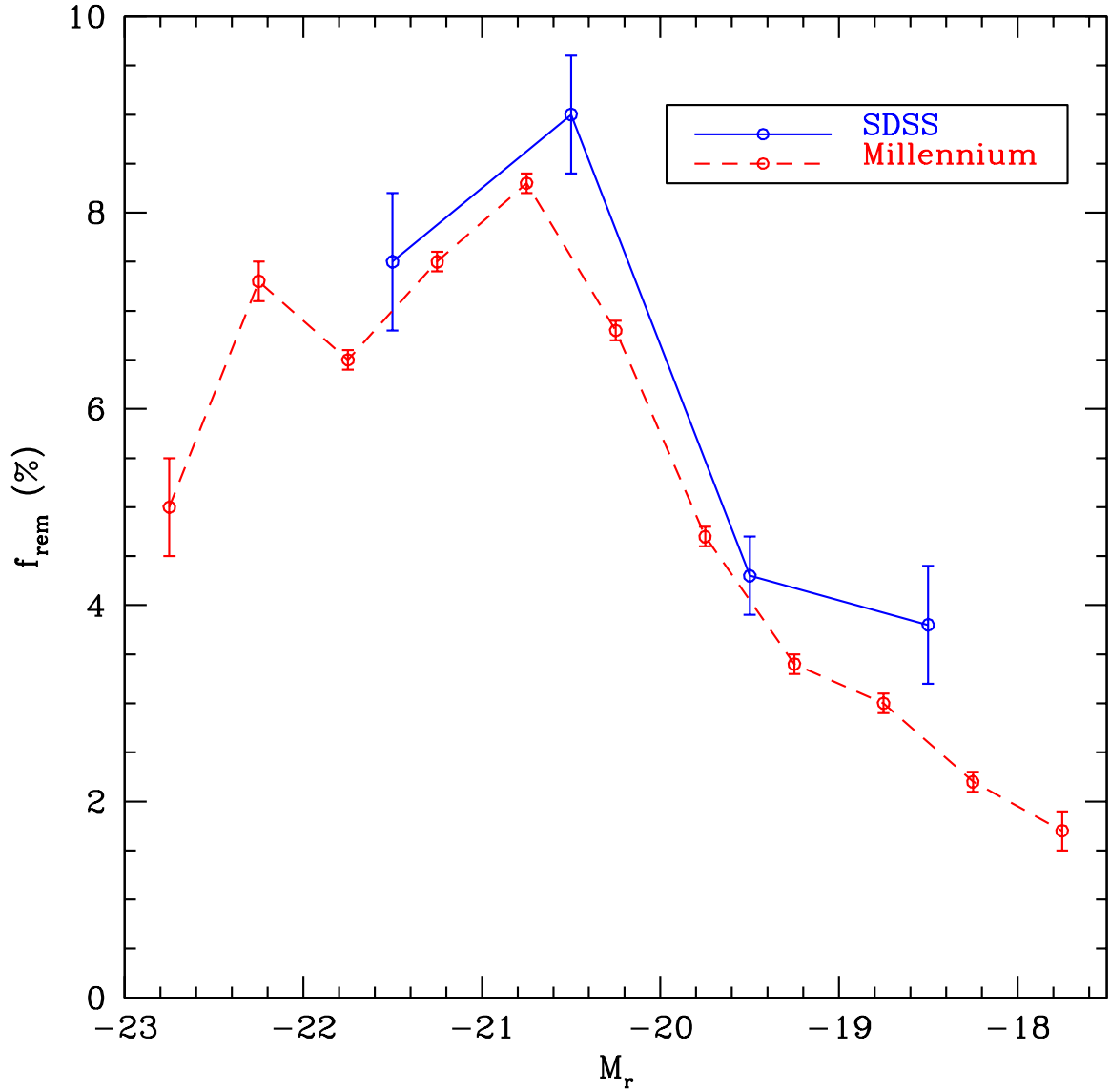}
\caption{The remnant fraction $f_{\rm rem}$, expressed as a percentage, 
is plotted versus $M_r$ for SDSS and for Millennium.
In both cases, the error bars are computed using only the uncertainties 
in $N_c$.  Given the assumptions that go into the calculation of $f_{\rm rem}$, 
the true uncertainties are likely to be considerably larger.
\label{figfrem}}
\end{figure}

Nevertheless, we present our no-evolution 
estimate of the luminosity dependent remnant 
fraction in Figure~\ref{figfrem} and in Tables~\ref{tabsdss} and \ref{tabmill}.
The SDSS remnant fraction 
peaks at 9\% for $M_r = -20.5$ (roughly $M^*$), and decreases to 
4\% over the range $-20 < M_r < -18$.  Similar trends are seen for 
Millennium, although the Millennium 
remnant fraction drops to $\sim 2\%$ at $M_r = -18$.

\subsection{The Luminosity Density of Close Companions}\label{lcv}

We demonstrated in Section~\ref{ncv} that our observed $n_c(M_r)$ implies that 
most major mergers occur between galaxies which are fainter than $L^*$.  
Of course, a major merger between two intrinsically faint galaxies 
affects much less stellar mass than a major merger of two luminous galaxies.
Therefore, in terms of the overall luminosity density of stellar mass that 
is participating in mergers, 
the number density of close companions does not tell the full story.
To address this question, we instead measure the luminosity 
of close companions 
per unit co-moving volume (hereafter $l_c(M_r)$), using luminosity  
as a proxy for stellar mass.  
We compute $l_c(M_r)$ as follows: 
\begin{equation}
l_c(M_r) = L_c(M_r) f_{\rm 3d}(M_r) n(M_r),
\end{equation} 
where $L_c(M_r)$ is the 
luminosity in close companions per galaxy \citep{ssrs2mr} as measured in 
redshift space, $f_{3d}(M_r)$ is as described in \S~\ref{f3d},  
and $n(M_r)$ is the co-moving number density of galaxies (see \S~\ref{ncv}).
When calculating luminosities, we use $M_r(\odot) = 4.64$ \citep{yasuda01}.  

\begin{figure}
\plotone{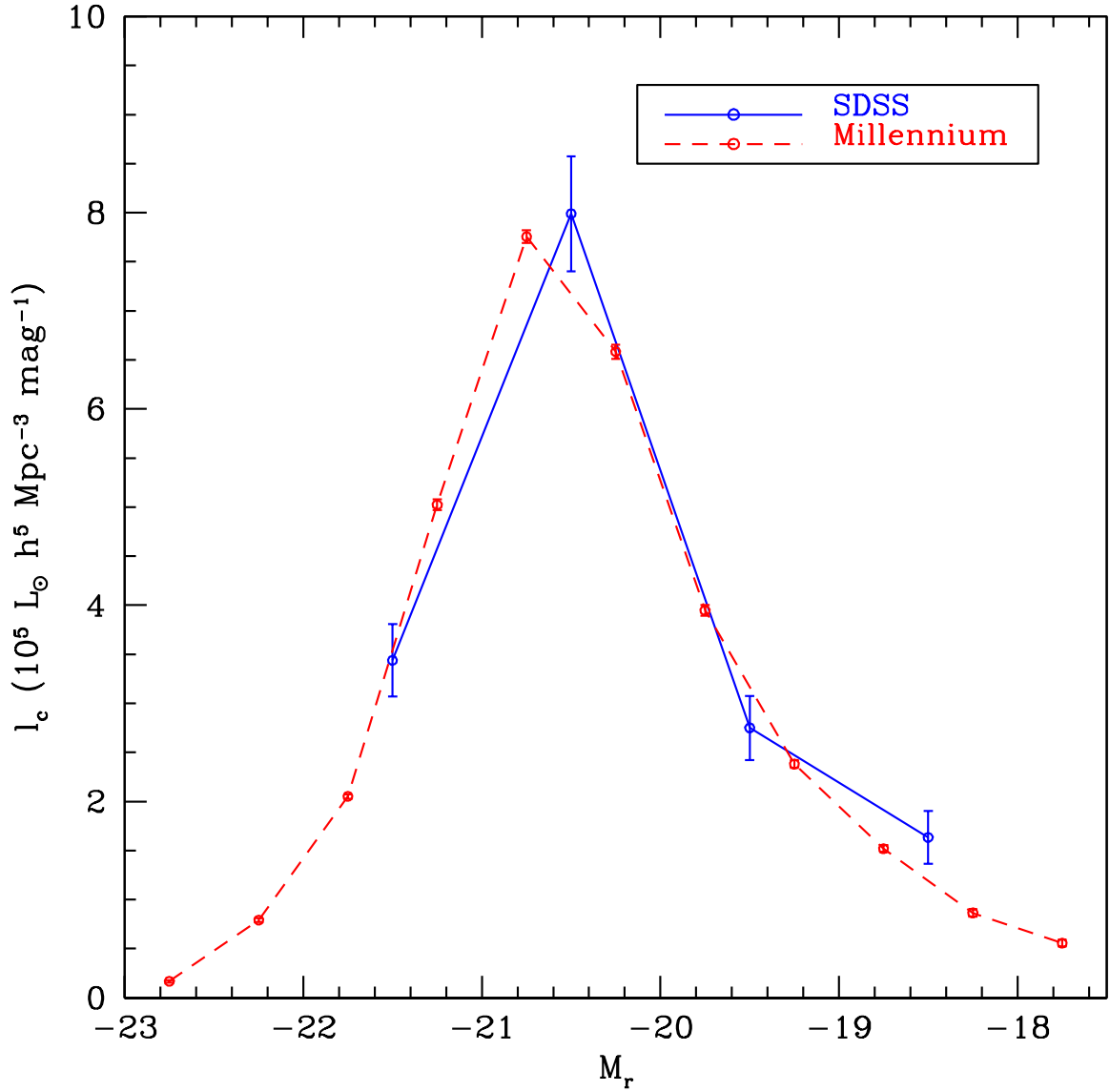}
\caption{The luminosity in close companions per unit co-moving volume ($l_c$) 
is plotted versus $M_r$ for SDSS and for Millennium.
Excellent agreement is seen between the two samples.  
$l_c(M_r)$ exhibits a clear peak at or very close to $M^*$.  
The decline towards brighter absolute magnitudes results primarily from the
low number density of such systems, whereas the decline at the faint end 
results from the decreasing luminosity of companions.  
In terms of stellar mass (as inferred from galaxy luminosities), 
this implies that $L^*$ is the most relevant regime for major mergers.
\label{figlc}}
\end{figure}

The resulting measurements for both SDSS and Millennium are given in 
Figure~\ref{figlc} and in Tables~\ref{tabsdss} and \ref{tabmill}.
Excellent agreement is again seen between SDSS and Millennium.  
Both surveys exhibit a clear peak in $l_c(M_r)$ at (or very close to) 
$M^*$, with a peak value of 
$l_c \sim 8 \times 10^5 \lsun h^5 {\rm Mpc}^{-3} {\rm mag}^{-1}$.  
The sharp decline towards the bright end is due to the lower 
number density of luminous host galaxies, while the decline towards the 
faint end is driven by the lower numbers and luminosities of companion galaxies.
This distribution clearly implies that galaxies which have luminosities 
close to $L^*$ are the most relevant in terms of the overall involvement 
of stellar mass in major mergers.

\section{Conclusions}\label{conclusions}

We have measured the number of close companions per galaxy ($N_c$) 
as a function of absolute magnitude for both the SDSS and 
the Millennium simulation.  For SDSS, we construct 
samples of host and companion galaxies, and correct for 
spectroscopic incompleteness.  For Millennium, we create a suite of 
mock redshift catalogs, averaging over different views of the 
\citet{croton06} cube.  
Using close pair criteria designed to identify imminent major 
mergers ($5 < r_p < 20~h^{-1}$ kpc, $\Delta v < 500$ km s$^{-1}$, and
$|\Delta M_r| < 0.753$), 
we find general agreement between the observations and simulations. 
In redshift space, $N_c \sim 0.02$ over the range $-22 < M_r < -18$.  
The flatness of this distribution indicates that $N_c(M_r)$ does not 
simply trace the number density of galaxies; instead, small scale 
luminosity-dependent clustering appears to counteract this effect.

Using three dimensional positions and velocities available from 
the Millennium simulations, we measure the degree to which the 
detected galaxy pairs are contaminated by projection effects, and find that 
the contamination is a strong function of absolute magnitude, 
rising from $45\%$ at $M_r \lesssim M^*$ to $\sim 83\%$ at $M_r = -18$.
We remove this contamination from both Millennium and SDSS pair statistics, 
yielding $N_c(M_r)$ measurements in real space.  These measurements 
indicate that 
galaxies with $M_r \lesssim M^*$ are the most likely to be undergoing 
major mergers at low redshift.  
However, by computing 
the number density of close companions ($n_c(M_r)$) in real space, 
we conclude that   
at least 90$\%$ of all major mergers occur between galaxies which 
are {\it fainter} than $M^*$.
With additional assumptions, we also 
estimate the galaxy and volume merger rates, which trace the real-space
$N_c(M_r)$ and $n_c(M_r)$ respectively.  We estimate that 
at least 8$\%$ of $L^*$ galaxies are likely to have undergone 
a major merger since $z=1$, while this remnant fraction appears to 
be $\sim 4$ times smaller for 0.1 $L^*$ galaxies.  Finally, our 
measurements of the luminosity density of close companions indicate 
a clear peak at $M^*$, implying that $L^*$ galaxies
dominate in terms of the overall amount of stellar mass involved 
in major mergers at low redshift.

Together, these results indicate that the low redshift merger rate depends 
strongly on luminosity (and presumably mass). 
This has a number of important implications.
For example, one would not expect the merger rates of massive galaxies 
(e.g., Masjedi et al. (2006)\nocite{masjedi06}) to agree with those 
of $L^*$ galaxies (e.g., De Propris et al. 2007\nocite{depropris07}). 
Also, the increase in projection effects for fainter galaxies indicates 
that samples of luminous galaxy pairs are more likely to provide 
bona fide merger candidates than samples of lower luminosity pairs.
This is relevant if one wishes to assess the impact of merging on the 
constituent galaxies (e.g., enhanced star formation or disturbed 
morphologies).
We also conclude that at low redshift, recent merging history is 
likely to be most important for galaxies which are relatively luminous.
Finally, given the clear peak in $l_c(M_r)$, it appears that 
galaxies which are much more or much less luminous than $L^*$ are unlikely 
to play an important role in the overall evolution of galaxies 
via major mergers.

\acknowledgments

We thank the anonymous referee for a thoughtful and constructive report, 
which led to significant enhancements of this paper.
We gratefully acknowledge the financial support of the Natural Sciences 
and Engineering Research Council (NSERC) of Canada, 
through a Discovery Grant to D. R. P. and
a USRA to J. E. A.  

Funding for the SDSS and SDSS-II has been provided by the Alfred P. Sloan Foundation, the Participating Institutions, the National Science Foundation, the U.S. Department of Energy, the National Aeronautics and Space Administration, the Japanese Monbukagakusho, the Max Planck Society, and the Higher Education Funding Council for England. The SDSS Web Site is http://www.sdss.org/.

The SDSS is managed by the Astrophysical Research Consortium for the Participating Institutions. The Participating Institutions are the American Museum of Natural History, Astrophysical Institute Potsdam, University of Basel, University of Cambridge, Case Western Reserve University, University of Chicago, Drexel University, Fermilab, the Institute for Advanced Study, the Japan Participation Group, Johns Hopkins University, the Joint Institute for Nuclear Astrophysics, the Kavli Institute for Particle Astrophysics and Cosmology, the Korean Scientist Group, the Chinese Academy of Sciences (LAMOST), Los Alamos National Laboratory, the Max-Planck-Institute for Astronomy (MPIA), the Max-Planck-Institute for Astrophysics (MPA), New Mexico State University, Ohio State University, University of Pittsburgh, University of Portsmouth, Princeton University, the United States Naval Observatory, and the University of Washington. 

The Millennium Run simulation used in this paper was carried out by the Virgo Supercomputing Consortium at the Computing Centre of the Max-Planck Society in Garching. The semi-analytic galaxy catalogue is publicly available at
http://www.mpa-garching.mpg.de\\/galform/agnpaper.

\clearpage

\begin{table}
\begin{center}
\caption{SDSS Host Galaxies
\label{tabhost}}
\begin{tabular}{cccccc}
\tableline\tableline
Host ObjID
&RA&Dec&$z$&$M_r$&Companion ObjID\\
\tableline
587724198275842071&10.610426&14.70509&0.05539&$-19.72$&587724198275842073\\
587724197207670886&23.645067&13.18706&0.06304&$-20.17$&587724197207670885\\
587724197207670885&23.646278&13.18500&0.06226&$-20.50$&587724197207670886\\
587731513153159200&42.731424&0.36690&0.04384&$-20.10$&587731513153159199\\
587731513153159199&42.739416&0.36940&0.04423&$-19.76$&587731513153159200\\
\tableline
\end{tabular}
\end{center}
\end{table}

\begin{table}
\begin{center}
\caption{SDSS Redshift Space $N_c(M_r)$ as a Function of Projected Pair Separation
\label{tabsdssnc}}
\begin{tabular}{ccccc}
\tableline\tableline
$r_p$ range&~$-22<M_r<-21$~&$-21<M_r<-20$&~$-20<M_r<-19$~&$-19<M_r<-18$\\
($h^{-1}$ kpc)&$(\bar{z}=0.107)$&$(\bar{z}=0.074)$&$(\bar{z}=0.048)$&$(\bar{z}=0.031)$\\
\tableline
5-20&0.0187 $\pm$ 0.0018&0.0238 $\pm$ 0.0016&0.0172 $\pm$ 0.0018&0.0226 $\pm$ 0.0034\\
20-40&0.0225 $\pm$ 0.0018&0.0275 $\pm$ 0.0016&0.0298 $\pm$ 0.0020&0.0284 $\pm$ 0.0026\\
40-60&0.0174 $\pm$ 0.0015&0.0332 $\pm$ 0.0015&0.0345 $\pm$ 0.0016&0.0329 $\pm$ 0.0024\\
60-80&0.0180 $\pm$ 0.0013&0.0346 $\pm$ 0.0011&0.0389 $\pm$ 0.0016&0.0398 $\pm$ 0.0026\\
80-100&0.0210 $\pm$ 0.0011&0.0360 $\pm$ 0.0011&0.0405 $\pm$ 0.0016&0.0470 $\pm$ 0.0028\\
\tableline
\end{tabular}
\end{center}
\end{table}

\begin{table}
\begin{center}
\small
\caption{SDSS Close Pair Statistics
\label{tabsdss}}
\begin{tabular}{ccccccc}
\tableline\tableline
$M_r$&$N_c$&$f_{3D}$&$N_c$&$n_c$
&~~~~~$f_{\rm rem}$~~~~~&$l_c$
\\
&~~~(redshift-space)~~~&($\%$)&~~~~~(real-space)~~~~~&&$(\%)$&\\
\tableline
-21.50 & 0.0187 $\pm$ 0.0018 & 55.3 $\pm$ 0.4& 0.0103 $\pm$ 0.0010& 1.52 $\pm$ 0.15& 7.5 $\pm$ 0.7 & 3.44 $\pm$ 0.37\\
-20.50 & 0.0238 $\pm$ 0.0016 & 52.5 $\pm$ 0.3& 0.0125 $\pm$ 0.0008& 6.93 $\pm$ 0.44& 9.0 $\pm$ 0.6 & 7.97 $\pm$ 0.58\\
-19.50 & 0.0172 $\pm$ 0.0018 & 33.8 $\pm$ 0.6& 0.0058 $\pm$ 0.0006& 5.78 $\pm$ 0.60& 4.3 $\pm$ 0.4 & 2.81 $\pm$ 0.33\\
-18.50 & 0.0226 $\pm$ 0.0034 & 22.8 $\pm$ 0.7& 0.0052 $\pm$ 0.0008& 6.78 $\pm$ 1.04& 3.8 $\pm$ 0.6 & 1.57 $\pm$ 0.26\\
\tableline
\end{tabular}
\large
\end{center}
\end{table}

\begin{table}
\begin{center}
\small
\caption{Millennium Close Pair Statistics
\label{tabmill}}
\begin{tabular}{ccccccc}
\tableline\tableline
$M_r$&$N_c$&$f_{3D}$&$N_c$&$n_c$
&~~~~~$f_{\rm rem}$~~~~~&$l_c$
\\
&~~~(redshift-space)~~~&($\%$)&~~~~~(real-space)~~~~~&&$(\%)$&\\
\tableline
-22.75 & 0.0142 $\pm$ 0.0010 & 45.9 $\pm$ 3.0& 0.0068 $\pm$ 0.0007& 0.03 $\pm$ 0.00& 5.0 $\pm$ 0.5 & 0.17 0.01\\
-22.25 & 0.0173 $\pm$ 0.0004 & 57.9 $\pm$ 1.2& 0.0101 $\pm$ 0.0003& 0.18 $\pm$ 0.01& 7.3 $\pm$ 0.2 & 0.79 0.02\\
-21.75 & 0.0163 $\pm$ 0.0002 & 54.9 $\pm$ 0.5& 0.0089 $\pm$ 0.0001& 0.71 $\pm$ 0.01& 6.5 $\pm$ 0.1 & 2.05 0.02\\
-21.25 & 0.0185 $\pm$ 0.0002 & 55.5 $\pm$ 0.6& 0.0103 $\pm$ 0.0001& 2.52 $\pm$ 0.03& 7.5 $\pm$ 0.1 & 5.02 0.06\\
-20.75 & 0.0207 $\pm$ 0.0002 & 55.5 $\pm$ 0.4& 0.0115 $\pm$ 0.0001& 5.57 $\pm$ 0.06& 8.3 $\pm$ 0.1 & 7.76 0.07\\
-20.25 & 0.0191 $\pm$ 0.0002 & 48.5 $\pm$ 0.6& 0.0093 $\pm$ 0.0002& 6.85 $\pm$ 0.12& 6.8 $\pm$ 0.1 & 6.58 0.07\\
-19.75 & 0.0176 $\pm$ 0.0002 & 36.6 $\pm$ 0.8& 0.0065 $\pm$ 0.0002& 6.39 $\pm$ 0.18& 4.7 $\pm$ 0.1 & 3.95 0.06\\
-19.25 & 0.0157 $\pm$ 0.0003 & 29.4 $\pm$ 0.7& 0.0046 $\pm$ 0.0001& 6.03 $\pm$ 0.16& 3.4 $\pm$ 0.1 & 2.38 0.04\\
-18.75 & 0.0162 $\pm$ 0.0004 & 25.0 $\pm$ 0.9& 0.0040 $\pm$ 0.0001& 6.09 $\pm$ 0.19& 3.0 $\pm$ 0.1 & 1.52 0.04\\
-18.25 & 0.0159 $\pm$ 0.0006 & 18.6 $\pm$ 1.2& 0.0029 $\pm$ 0.0002& 5.44 $\pm$ 0.35& 2.2 $\pm$ 0.1 & 0.86 0.04\\
-17.75 & 0.0149 $\pm$ 0.0009 & 15.4 $\pm$ 1.9& 0.0023 $\pm$ 0.0003& 5.41 $\pm$ 0.71& 1.7 $\pm$ 0.2 & 0.56 0.04\\
\tableline
\end{tabular}
\large
\end{center}
\end{table}

\clearpage

\end{document}